\documentclass[a4paper,11pt]{article}
\pdfoutput=1 % if your are submitting a pdflatex (i.e. if you have
\usepackage{jheppub} % for details on the use of the package, please
\usepackage{physics}
\usepackage[T1]{fontenc} % if needed
\usepackage{appendix}
\usepackage{amsmath, amssymb, graphicx}
\usepackage{amsfonts}
\usepackage{stmaryrd}
\usepackage{mathtools}

\usepackage{amsthm, color, xcolor, hyperref, slashed, subfigure, ytableau}
\ytableausetup{boxsize=0.8em}

\def \bal#1\eal  {\begin{align} #1 \end{align}}
\def \bga#1\ega  {\begin{gather} #1 \end{gather}}
\def\({\left(}
\def\){\right)}
\def\[{\left[}
\def\]{\right]}
\def\<{\left\langle}
\def\>{\right\rangle}
\def\d{\mathrm{d}}
\def\i{\int}

\newcommand{\yt}[1] {\begin{ytableau} #1\end{ytableau} }

\newcommand{\f}[2]{\frac{#1}{#2}}
\newcommand{\bim} {\begin{itemize}[noitemsep]}

\newcommand{\eim}{\end{itemize}}
\newcommand{\beq} {\begin{equation}}
\newcommand{\eeq} {\end{equation}}
\newcommand{\bc}{\begin{center}}
\newcommand{\ec}{\end{center}}

\newcommand{\nn} {\nonumber\\}

\newcommand{\ie}{~~~~{\rm ie,}~~~~}

\newcommand{\pr}{\prime}

\newcommand{\Hom}{{\rm Hom}}

\title{Multi-Matrix Correlators and Localization}

\author[a,b]{Adolfo Holguin,}
\author[a]{Shannon Wang,}
\author[a]{Zi-Yue Wang}

\affiliation[a]{Department of Physics, University of California, Santa Barbara, CA 93106, USA}

\affiliation[b]{School of Mathematics and Hamilton Mathematics Institute, 17 Westland Row Trinity College Dublin, Dublin 2, Ireland}

\emailAdd{holguina@tcd.ie}
\emailAdd{shannonwang@physics.ucsb.edu}
\emailAdd{zi-yue@physics.ucsb.edu}

\abstract{We study generating functions of $\frac{1}{4}$-BPS states in $\mathcal{N}=4$ super Yang-Mills at finite $N$ by attempting to generalize the Harish-Chandra-Itzykson-Zuber integral to multiple commuting matrices. This allows us to compute the overlaps of two or more generating functions; such calculations arise in the computation of two-point correlators in the free-field limit. We discuss the four-matrix HCIZ integral in the $U(2)$ context and lay out a prescription for finding a more general formula for $N>2$. We then discuss its connections with the restricted Schur polynomial operator basis. Our results generalize readily to arbitrary numbers of matrices, opening up the opportunity to study more generic BPS operators. }

\begin{document} 
\maketitle
\flushbottom

\section{Introduction}
\label{sec:first}

Large operators in large $N$ gauge theories are an important subject of study with relevance to nuclear physics, theories of quantum gravity and strings. Although there has been enormous success in computing the spectrum of anomalous dimensions of light operators in models such as maximally supersymmetric Yang-Mills theory in the planar limit, very little is known about how to tackle generic operators whose dimensions can scale with a power of $N$. This is an interesting problem for holography \cite{Maldacena:1997re} and for understanding the structure of conformal field theories more generally \cite{Hellerman:2015nra}. One of the difficulties one faces when trying to address these types of problems is that the intuitions from the planar limit are often unjustified for large operators; one must sum over both planar and non-planar diagrams and it is not a priori clear which diagrams dominate in the large $N$ limit. A promising approach is to replace single and multi- trace operators with a different basis that is better behaved at finite $N$ \cite{Corley:2001zk}, and then perform a systematic expansion around protected states in the large $N$ limit. In the case of maximally supersymmetric Yang-Mills theory, this has been implemented at finite $N$ \cite{deMelloKoch:2007rqf, Bhattacharyya:2008rb, Berenstein:2005fa, Carlson:2011hy}. Even though the expressions found through these techniques at finite $N$ are quite explicit, it is usually difficult to take the large $N$ limit of such quantities.

More recently, there have been works showing that certain generating functions can be used to perform computations in the free-field theory limit \cite{Jiang:2019xdz, Berenstein:2022srd, Chen:2019gsb, Holguin:2022drf, Lin:2022wdr}. This technique has been succesfully implemented in the computation of three-point correlators involving large operators made out of a single matrix field \cite{Yang:2021kot, Holguin:2022zii, Holguin:2023orq}, as in the half-BPS sector of $\mathcal{N}=4$ SYM \cite{Berenstein:2004kk, Lin:2004nb}, where the dual gravitational description is explicitly realized from the gauge theory. An explicit mapping between BPS states made out of more than one matrix and asymptotically $AdS_5\times S^5$ geometries is still lacking, though a compelling description in terms of \textit{bubbling geometries} seems to exist \cite{Berenstein:2005aa, Chen:2007du, lin2010studies}. The study of generating functions for multi-matrix correlators was outlined in \cite{Chen:2019gsb, Carlson:2022dot} for certain classes of operators, and more generally in \cite{Berenstein:2022srd}. Our goal is to elucidate some of the details regarding the generating functions of $\frac{1}{4}$ and $\frac{1}{8}$- BPS operators in $\mathcal{N}=4$ SYM. We do this by proposing a fixed-point formula for the overlap of generic coherent state generating functions; this gives us an integral formula that generalizes the Harish-Chandra-Itzykson-Zuber (HCIZ) formula to multiple pairs of matrices. Integrals of this type appear naturally in the study of multi-matrix models of commuting random matrices. 

This paper is structured as follows. In section \ref{sec:second}, we review the generating function techniques, focusing on the case of $\frac{1}{4}$ BPS operators in $\mathcal{N}=4$ SYM. We argue that the form of these operators is protected, so we can restrict to eigenstates of the one-loop dilatation operator. We then evaluate the norm of the generating function for the $U(2)$ theory by explicit integration to motivate our fixed point formula for general $N$. Finally, we give a prescription for extending the HCIZ formula to the multiple-matrix model using the heat kernel method as outlined in \cite{itzykson1980planar}; we will discuss our results for $U(2)$ and $U(3)$ and the insights we may glean from them to extrapolate a general formula for $U(N)$. In section \ref{sec:third}, we connect our results to the construction of restricted Schur polynomials and outline how to generalize to operators associated with Young diagrams with arbitrary number of rows or columns. We will briefly discuss our attempts to arrive at a general formula via the character expansion method. Finally, we conclude with some future directions.

\section{Multi-matrix Generating Functions}
\label{sec:second}

We are interested in studying operators in gauge theories that are made out of more than one matrix-valued scalar field. In particular, we will work with $\frac{1}{4}$-BPS operators in $U(N)$ $\mathcal{N}=4$ SYM on the cylinder $\mathbb{R}\times S^3$. At weak coupling, these operators can be built out of symmetrized products of two of the three complex scalar fields of the theory $X,Y$. Generalizing to more than two matrices is straightforward. This class of operators transforms in the $[p,q,p]$ representations of the $SU(4)_R$ symmetry, and the operators are generically of multi-trace form. We will concentrate on scalar primary states at an equal time slice for simplicity. Unlike $\frac{1}{2}$-BPS operators, which can be built explicitly in the free theory, $\frac{1}{4}$-BPS operators of the interacting theory are different from those of the free theory. The lifting of states due to non zero gauge coupling can be treated pertubatively and the loop corrections to dilatation operators annihilate operators that are made out of symmetric products of $X$ and $Y$. This problem was studied in detail for small operators in \cite{Ryzhov:2001bp}, but for generic large operators, explicit constructions in terms multi-traces are cumbersome. An alternative expansion in terms of characters was introduced in \cite{deMelloKoch:2007rqf}, which the authors call the \textit{restricted Schur polynomial basis}. This basis is convenient for dealing with the mixing between the different trace structures since it diagonalizes the matrix of two point functions for all values of $N$.  

\subsection{Generating $\frac{1}{4}$ BPS States}
\label{sec:secondsubone}

Yet another way of generating $\frac{1}{4}$-BPS states can be found by studying operators of the form:
\begin{equation}
\label{eqn:twomatrixstate}
    \ket{\Lambda_X, \Lambda_Y}= \frac{1}{\text{Vol}\left[U(N)\right]} \int dU\; \exp\left(\Tr\left[U X U^\dagger \Lambda_X +UY U^\dagger \Lambda_Y\right]\right) \ket{0}.
\end{equation}

If we insist that the coherent state parameters $\Lambda_X$ and $\Lambda_Y$ commute, $\ket{\Lambda_X, \Lambda_Y}$ is annihilated by the one-loop dilatation operator; it was shown in \cite{Lin:2022wdr} that this persists to two-loop order. In \cite{Grant:2008sk}, it was conjectured that the space of BPS states in $\mathcal{N}=4$ SYM is given by the kernel of the one-loop dilatation operator at all values of the coupling; we will take this as a working assumption and work with the set of states annihilated by the Beisert one-loop dilatation operator:
\begin{equation}
    \hat{D}_2^{SU(2)}= g^2 \Tr\left[[X,Y][\partial_X, \partial_Y]\right].
\end{equation}

Because the  states \eqref{eqn:twomatrixstate} are coherent states of $\bar{X}, \bar{Y}$ \cite{Berenstein:2022srd}, they form an overcomplete basis of states for any value of $N$. This has many computational advantages, mostly due to the fact that taking the large $N$ limit is very straightforward, but translating back into a complete orthogonal basis of operators can be complicated. This may be solved by computing the norm of the coherent states. By exploiting the Campbell-Hausdorff formula, we arrive at an integral of the form:
\begin{equation}
\label{eqn:twopointfunction}
    \bra{\bar{\Lambda}_X, \bar{\Lambda}_Y}\ket{\Lambda_X, \Lambda_Y}=  \frac{1}{\text{Vol}\left[U(N)\right]} \int dU\; \exp\left(\Tr\left[U \bar{\Lambda}_X U^\dagger \Lambda_X +U\Bar{\Lambda}_Y U^\dagger \Lambda_Y\right]\right).
\end{equation}

Since we can in principle expand \eqref{eqn:twomatrixstate} in terms of an orthonormal basis, we may use this overlap to determine the coefficients relating the multi-trace basis of operators to an orthogonal basis by expanding in a series and matching the coefficients as done in \cite{Berenstein:2022srd}. The precise tool relating the multi-trace basis operators and the character expansion in this case is the Weingarten calculus \cite{collins2021weingarten}; an example illustrating this technique can be found in \cite{Diaz:2013gja}. The main obstacle we face is evaluating the integral \eqref{eqn:twopointfunction} for generic coherent state parameters. To our knowledge, these types of integrals have not been studied before, and a closed form expression for them is needed. Our main goal will be to evaluate this class of integrals for any value of $N$. Although we only explicitly study the case of $U(N)$ integrals, the methods should apply generally and should generalize to $SO(N)$ and $Sp(N)$ groups as well as to quivers. These types of integrals are also a natural object to study in the context of matrix models, since they arise in the study of multi-matrix models of commuting matrices.

\subsection{The Four-Matrix Model in $SU(2)$}
\label{secondsubtwo}

Before proceeding to the case of general $N$, we will study the following integral
\begin{equation}
\label{eqn:hcizsu2}
   I_2= \int_{SU(2)} dU \; e^{\Tr\left[U A U^\dagger \bar{A}+U B U^\dagger \bar{B} \right]}
\end{equation}
for commuting matrices $A,B, \bar{A}, \bar{B}$. We will first approximate $I_2$ by a saddle point approximation; the critical points of the function in the exponential are given by the solutions to the equations 
\begin{equation}
    [A, U^\dagger \bar{A} U]+  [B, U^\dagger \bar{B} U]=0.
\end{equation}
For generic enough matrices, this is only satisfied if each of the two terms vanishes individually
\begin{equation}
\begin{aligned}
       [A, U^\dagger \bar{A} U]=[B, U^\dagger \bar{B} U]=0.
\end{aligned}
\end{equation}

The only problematic cases occur when a subset of the eigenvalues of $B$ is a permutation of a subset of eigenvalues of $-A$. From here on, we assume that the eigenvalues are generic enough that this does not happen. This means that, generically, the saddle points are labelled by permutation matrices $U_\pi$. We are then left with a Gaussian integral around each of the saddle points, which can be evaluated easily; this results in a "one-loop determinant" factor given by:
\begin{equation}
    D_2(a,\bar{a}, b, \bar{b})=\left(a_1 - a_2\right)\left(\bar{a}_1 - \bar{a}_2\right) + \left(b_1 - b_2\right)\left(\bar{b}_1 - \bar{b}_2\right)
\end{equation}

This gives an approximate value for the integral (up to a convention dependent normalization factor):
\begin{equation}
    I_2\simeq \frac{e^{a_1 \bar{a}_1+a_2 \bar{a}_2+b_1 \bar{b}_1+b_2 \bar{b}_2}-e^{a_1 \bar{a}_2+a_2 \bar{a}_1+b_1 \bar{b}_2+b_2 \bar{b}_1}}{\left(a_1 - a_2\right)\left(\bar{a}_1 - \bar{a}_2\right) + \left(b_1 - b_2\right)\left(\bar{b}_1 - \bar{b}_2\right)}.
\end{equation}
At first sight, it is not clear that this approximation is reliable, since there is no large parameter in the exponential. To gain more intuition, we evaluate $I_2$ through an explicit computation. 

First, we must parameterize our unitary matrix $U$; then, we need to compute the Haar measure. We start with the following matrices:
\begin{equation}
\begin{aligned}
A &= \begin{pmatrix}a_1 & 0\\0 & a_2\end{pmatrix},\;B = \begin{pmatrix}b_1 & 0\\0 & b_2\end{pmatrix}\\
\bar{A} &= \begin{pmatrix}\bar{a}_1 & 0\\0 & \bar{a}_2\end{pmatrix},\;\bar{B} = \begin{pmatrix}\bar{b}_1 & 0\\0 & \bar{b}_2\end{pmatrix}
\end{aligned}
\end{equation}

We then seek to parametrize our unitary matrix. We know that any arbitrary $SU(2)$ matrix must meet the following conditions:
\begin{equation}
\label{eqn:su2}
SU(2) = \left\{\begin{pmatrix}a & b\\ -b^* & a^*\end{pmatrix} \in \mathbb{C}^{2\times 2} \; \Big| \; |a|^2 + |b|^2 = 1\right\}
\end{equation}

For ease of computation, we choose to parameterize $U$ with Euler angles: 
\begin{equation}
\label{eqn:umatrix}
U = \begin{pmatrix}e^{-i\frac{\gamma+\alpha}{2}}\cos\frac{\theta}{2} & -e^{i\frac{\gamma-\alpha}{2}}\sin\frac{\theta}{2} \\ e^{-i\frac{\gamma-\alpha}{2}}\sin\frac{\theta}{2} & e^{i\frac{\gamma+\alpha}{2}}\cos\frac{\theta}{2}\end{pmatrix}
\end{equation}

We seek to rewrite the Haar measure $dU$ in terms of $J(\theta,\gamma,\alpha)d\theta d\gamma d\alpha$, where $J(\theta,\gamma,\alpha)$ is the Jacobian. We may do so by computing the inverse of the unitary matrix and multiplying it by its partial derivatives with respect to the Euler angles. We start by finding the inverse of $U$:

\begin{equation}
\label{eqn:uinverse}
U^{-1} = \begin{pmatrix}e^{i\frac{\gamma+\alpha}{2}}\cos\frac{\theta}{2} & e^{i\frac{\gamma-\alpha}{2}}\sin\frac{\theta}{2} \\ -e^{-i\frac{\gamma-\alpha}{2}}\sin\frac{\theta}{2} & e^{-i\frac{\gamma+\alpha}{2}}\cos\frac{\theta}{2}\end{pmatrix}
\end{equation}

Then we calculate the partial derivatives with respect to $\gamma$, $\alpha$, and $\theta$ and multiply by the inverse. We obtain:
\begin{equation}
\begin{aligned}
U^{-1}\frac{\partial U}{\partial\gamma} &= \begin{pmatrix}-\frac{i}{2} & 0\\ 0 & \frac{i}{2}\end{pmatrix}\\
U^{-1}\frac{\partial U}{\partial\alpha} &= \begin{pmatrix}-\frac{i}{2}\cos\theta & \frac{i}{2}e^{i\gamma}\sin\theta\\ \frac{i}{2}e^{-i\gamma}\sin\theta & \frac{i}{2}\cos\theta\end{pmatrix}\\
U^{-1}\frac{\partial U}{\partial\theta} &= \begin{pmatrix}0 & -\frac{1}{2}e^{i\gamma}\\ \frac{1}{2}e^{-i\gamma} & 0\end{pmatrix}
\end{aligned}
\end{equation}

We calculate the Jacobian matrix using the following basis $\epsilon_1 = \begin{pmatrix}i & 0\\0 & -i\end{pmatrix}$, $\epsilon_2 = \begin{pmatrix}0 & ie^{i\gamma}\\ie^{-i\gamma} & 0\end{pmatrix}$, and $\epsilon_3 = \begin{pmatrix}0 & -e^{i\gamma}\\e^{-i\gamma} & 0\end{pmatrix}$:
\begin{equation}
\mathcal{J} = \begin{pmatrix}-\frac{1}{2} & -\frac{1}{2}\cos\theta & 0\\0 & \frac{1}{2}\sin\theta & 0 \\0 & 0 & \frac{1}{2}\end{pmatrix}  
\end{equation}

The Jacobian $J(\theta,\gamma,\alpha)$ we seek is the determinant of $\mathcal{J}$:

\begin{equation}
    \det(J) = \frac{1}{8}|\sin\theta|
\end{equation}

We see that it is only dependent on $\theta$. Our integral becomes:
\begin{equation}
\begin{aligned}
I_2 &= \frac{1}{8}\int_0^{\pi}d\theta\int_0^{4\pi}\frac{d\gamma}{4\pi}\int_0^{2\pi}\frac{d\alpha}{2\pi}|\sin\theta|e^{\Tr\left[\bar{A}UAU^\dagger + \bar{B}UBU^\dagger\right]} \\
&= \frac{1}{8}\int_0^{\pi}d\theta|\sin\theta|e^{\frac{1}{2}\left(\left(a_1 + a_2\right)\left(\bar{a}_1 + \bar{a}_2\right) + \left(b_1 + b_2\right)\left(\bar{b}_1 + \bar{b}_2\right) + \left(\left(a_1 - a_2\right)\left(\bar{a}_1 - \bar{a}_2\right) + \left(b_1 - b_2\right)\left(\bar{b}_1 - \bar{b}_2\right)\right)\cos\theta\right)}
\end{aligned}
\end{equation}

Our critical points are $\theta = 0$ and $\theta = \pi$, so we can remove the absolute value bars. Then we evaluate our integral:
\begin{equation}
\label{eqn:su2fixedpointformula}
\begin{aligned}
    I_2 &= \frac{1}{8}\int_0^{\pi}d\theta\sin\theta e^{\frac{1}{2}\left(\left(a_1 + a_2\right)\left(\bar{a}_1 + \bar{a}_2\right) + \left(b_1 + b_2\right)\left(\bar{b}_1 + \bar{b}_2\right) + \left(\left(a_1 - a_2\right)\left(\bar{a}_1 - \bar{a}_2\right) + \left(b_1 - b_2\right)\left(\bar{b}_1 - \bar{b}_2\right)\right)\cos\theta\right)}\\
    &= \frac{e^{a_1\bar{a}_1 + a_2\bar{a}_2 + b_1\bar{b}_1 + b_2\bar{b}_2} - e^{\bar{a}_1a_2 + a_1\bar{a}_2+\bar{b}_1b_2 + b_1\bar{b}_2}}{4\left(\left(a_1 - a_2\right)\left(\bar{a}_1 - \bar{a}_2\right) + \left(b_1 - b_2\right)\left(\bar{b}_1 - \bar{b}_2\right)\right)}
\end{aligned}
\end{equation}

This is precisely the same result that the saddle point approximation yields. From the intermediate steps, it is clear that there are never any terms that mix the eigenvalues of $A$ and $B$; if we set either $A=0$ or $B=0$, we immediately recover the HCIZ formula for $U(2)$. 

\subsection{Proof of Localization for $U(2)$}
\label{sec:secondsubthree}

One important issue to understand is why the integral $I_2$ has an exact saddle point approximation, while the naive saddle point approximation for $I_{N}$ fails to be exact for $N>2$. The idea is to try to follow the proof \cite{itzykson1980planar} for the HCIZ integral and see exactly how the analysis differs for multi-matrix models. One key observation is that the integral $I_N(A,B, \bar{A}, \bar{B})$ is an eigenfunction of a holomorphic Laplacian:
\begin{equation}
  - \left[ \frac{\partial}{\partial A_{ij}}\frac{\partial}{\partial A_{ji}}+ \frac{\partial}{\partial B_{ij}}\frac{\partial}{\partial B_{ji}}\right] I_N(A,B, \bar{A}, \bar{B})= \Tr[\bar{A}^2 + \bar{B}^2] I_N(A,B, \bar{A}, \bar{B}).
\end{equation}
Before continuing, it is worthwile to explain what we mean by holomorphic in this context, and why this is important. One way in which the integral $I_N$ appears is as the Jacobian factor for a Gaussian matrix integral over a pair of commuting normal matrices
\begin{equation}
\begin{aligned}
 \mathcal{Z}&=  \int_{[A,\bar{A}]=0} [dA d\bar{A}]\int_{[B, \bar{B}]=0}[dB d\bar{B}]\; \exp\bigg\{ \Tr[A \bar{A}]+\Tr[B \bar{B}]\bigg\} \;\delta\big([A,B]\big)\\
 &=\int d\mu(a, \bar{a}, b, \bar{b}) \;I_N(A, B,\bar{A}, \bar{B})
 \end{aligned}
\end{equation}
As it stands, this expression is formal unless we specify a contour of integration for the eigenvalues of $A, \bar{A}$ and $B, \bar{B}$. A choice of contour corresponds to a choice of polarization in the space of eigenvalues; this makes the eigenvalues of $A$ and $\bar{A}$ canonically conjugate. This is quite natural from the interpretation of the integral as the norm of a coherent state in matrix quantum mechanics, where the collective coordinates $a_i$ and $b_i$ are holomorphic phase space coordinates. The barred coordinates are then conjugate momentum variables. Thus, the correct Laplacian operator has to be constructed from the metric of the space of commuting normal matrices. This is exactly the quantization discussed in \cite{Berenstein:2004kk}.

We rescale the matrices by a constant factor $t$; the resulting equation implies that the integral is related to a holomorphic heat kernel on the space of commuting matrices:
\begin{equation}
    K_t(a, \bar{a}, b, \bar{b})=t^{-N} \int dU \exp\left\{-\frac{1}{t}\Tr[U AU^\dagger \bar{A}+U BU^\dagger \bar{B}] \right\}
\end{equation}
As we take $t$ to zero, the integral will be very well approximated by the saddle point approximation. We see that the integral approaches a delta function; we can use the kernel itself to propagate this initial condition to a finite $t$. This would imply that the integral comes from a sum over the real saddle points of the integral. If the kernel is a plane wave, then the integral localizes, which is to say that the steepest descent contour gives exactly a Gaussian integral centered around each saddle point. This occurs if the heat equation for the kernel corresponds to a Sch\"odinger equation for an integrable system, since we can in principle change the variables into a set of action-angle variables, where the wavefunction is a plane wave. Whenever the kernel cannot be written this way, true localization fails, and instead, the integral is given by a sum over thimbles, with the kernel giving a parametrization of the integration contour. \\

Now we review some coordinate transformations for the Laplacian in the space of normal matrices. Given that the squared distance of two $n\times n$ normal matrices $A$ and $A'$ is $d(A, A') = \Tr|A-A'|^2$ and invariant under unitary transformations $A \rightarrow UAU^\dagger$, the metric is:
\begin{equation}
\label{eqn:metrictensor}
ds^2 = \sum_{ij}|dA_{ij}|^2.
\end{equation}
 We know that the Laplace-Beltrami operator is:
\begin{equation}
\nabla^2 = \frac{1}{\sqrt{g}}\partial_i\left(g^{ij}\sqrt{g}\partial_j\right)
\end{equation}

We now perform a coordinate transformation $A = U\Omega aU^\dagger$ to rewrite the matrix in terms of its $n$ eigenvalues $a_i$ and $n(n-1)$ angular variables $\theta_\alpha$. Setting $dH = iU^\dagger dU$, we may rewrite our invariant distance as:
\begin{equation}
\begin{aligned}
ds^2 &= g_{i\bar{j}}da_i d\bar{a}_j + g_{\alpha\beta}d\theta^\alpha d\theta^\beta\\
&= \sum_i|da_i|^2 + \sum_{i,j}|a_i-a_j|^2dH_{ij}dH_{ji},
\end{aligned}
\end{equation}
where we have defined:
\begin{equation}
g_{\alpha\beta} = 2\sum\limits_{i<j}|a_i-a_j|^2\text{Re}(\partial_\alpha H_{ij})(\partial_\beta H^*_{ij}) = (VDV^\dagger)_{\alpha\beta},
\end{equation}
where $V_{\alpha,ij} = \partial_\alpha H_{ij}$ and $D$ is a diagonal matrix with elements $|a_i-a_j|^2$. The square root of the metric tensor's determinant is then:
\begin{equation}
\sqrt{g} = |\Delta|^4|\det V|,
\end{equation}
where $\Delta$ is the Vandermonde determinant of the eigenvalues $a_i$. Because the new metric tensor is block diagonal with an eigenvalue sector and a unitary sector, its inverse is block diagonal as well with an eigenvalue sector and a unitary sector, and thus the Laplacian may be separated into two operators, one for each sector. In its entirety, the Laplacian is:
\begin{equation}
\label{eqn:laplacian}
\nabla^2_A = \frac{1}{|\Delta|^4}\sum_i\frac{\partial}{\partial a_i}|\Delta|^4\frac{\partial}{\partial a_i} + ]\frac{1}{|\Delta|^2}\frac{1}{\det|V|}\sum_\alpha\frac{\partial}{\partial \theta_\alpha}(V^{-1})^*_{\alpha,ij}|\det V|\sum_\beta(V^{-1})_{ij,\beta}\frac{\partial}{\partial \theta_\beta}
\end{equation}

We now consider the space of two $N\times N$ commuting normal matrices $A$ and $B$. After diagonalization the metric for this space becomes:

\begin{equation}
 ds^2 = |da_i|^2 + |db_i|^2 + \sum_{i,j}\left(|a_i-a_j|^2+|b_i-b_j|^2\right)dH_{ij}dH_{ji},    
\end{equation}

Then the square root of the metric tensor's determinant becomes:

\begin{equation}
\sqrt{g} = \prod_{i<j}\left(|a_i-a_j|^2+|b_i-b_j|^2\right)^2|\det V|= \mu^2 |\det V|,
\end{equation}
and we rewrite $\prod_{i<j}\left(|a_i-a_j|^2+|b_i-b_j|^2\right)$ as $\mu$. 

We may rewrite the holomorphic Laplacian as \cite{Berenstein:2004kk}:

\begin{equation}
\begin{aligned}
\nabla^2_{A,B} &= \frac{1}{\mu^2}\left[\sum_k\frac{\partial}{\partial a_k}\mu^2\frac{\partial}{\partial a_k} + \sum_k\frac{\partial}{\partial b_k}\mu^2\frac{\partial}{\partial b_k}\right]\\
&+ \sum_{i<j}\frac{1}{\mu^2}\frac{1}{\det|V|}\sum_\alpha\frac{\partial}{\partial \theta_\alpha}(V^{-1})^*_{\alpha,ij}|\det V|\sum_\beta(V^{-1})_{ij,\beta}\frac{\partial}{\partial \theta_\beta}\\
\end{aligned}
\end{equation}
Notice that all of the eigenvalue dependance is on the first two terms and because the integral averages over the angular variables of $A$ and $B$ it is annihilated by the last term, so we will omit it from now on. So now our problem is reduced to finding eigenfunctions for this operator. As is common in matrix quantum mechanics, one can often reabsorb the measure factor $\mu$ into the definition of the eigenfuntion, so we will express $I_N(A,B, \bar{A}, \bar{B})$ in terms of an auxiliary function $\Psi_N(A,B, \bar{A}, \bar{B})$:
\begin{equation}
    I_N(A,B, \bar{A}, \bar{B})= \mu \,\Psi_N(A,B, \bar{A}, \bar{B}).
\end{equation}
After this rescaling, the Laplacian operator becomes a sum of two terms, one being the flat space Laplacian and the other an effective potential:
\begin{equation}
   \nabla^2_{A,B} \, I_N(A,B, \bar{A}, \bar{B})= \frac{1}{\mu}\left[\nabla_{a}^2 + \nabla_{b}^2\right] \Psi_N - \frac{1}{\mu^2}\left(\left[\nabla_{a}^2 + \nabla_{b}^2\right]\mu\right) \Psi_N= \frac{\lambda \,\Psi_N}{\mu}.
\end{equation}
So far our discussion applies to general rank of matrices. Focusing on $N=2$, we can easily check that the potential term vanishes. This is because $\mu$ is linear in $a_i$ and $b_i$. In this case, the problem reduces to finding eigenfunctions for the Laplace operator in flat space:
\begin{equation}
    \left[\nabla_{a}^2 + \nabla_{b}^2\right] \Psi_2= \lambda \Psi_2.
\end{equation}
The solutions to this equation are plane waves: 
\begin{equation}
    \Psi_2\sim\prod_i e^{a_i \bar{a}_i+ b_i \bar{b}_i}.
\end{equation}
This ansatz does not respect the symmetry properties of the integral under simultaneous permutations of $a_i$ and $b_i$, so the correct solution is a symmetrized sum of plane waves:
\begin{equation}
     \Psi_2= \mathcal{C}_2 \;\frac{1}{2!}\sum_{\pi \in S_2} (-1)^{\pi}\,\prod_i e^{a_i \bar{a}_{\pi(i)}+ b_i \bar{b}_{\pi(i)}}.
\end{equation}
After dividing by the measure factor, we reproduce the expected answer. At this point, it becomes clear that the heat kernel proof works for the $SU(2)$ integral, since the kernel is Gaussian and the saddle point approximation as $t\rightarrow 0$ can be propagated forward to obtain the integral for finite $t$. Intuitively, we should be able to localize the integral much like the single matrix case, because the wavefunction $\Psi_2$ is an eigenfuction of an integrable (free) Hamiltonian. In the case where $B=\bar{B}=0$, the measure factor $\mu$ reduces to a Vandermonde determinant, which is also annihilated by the flat space Laplacian; thus we see that the usual HCIZ integral is associated with a wavefunction of a free fermion or boson. But this is not the case for $B\neq0$ and $N>2$, since the potential term does not vanish. Note that this does not necessarily mean that the integral is not localizable. An example that comes to mind are the integrals of the Harish-Chandra type for the symplectic groups, which are associated with the wavefunctions of integrable Calogero models. While these integrals are known to localize by the heat kernel methods \cite{McSwiggen_2018}, in this case, it was noted that the naive localization argument nevertheless still fails \cite{Brezin_2003}, and that one must include additional instanton solutions to the WKB approximation. 

Returning to our guess for $\Psi_N$, it becomes clear that a closed form for $ I_N(A,B, \bar{A}, \bar{B})$ must include $\mu$ as its denominator. This is because $\mu$ is the natural integration measure for the eigenvalues $a_i, b_i$. The fact that the eigenvalue is $\Tr[\bar{A}^2+ \bar{B}^2]$ also suggests that the denominator should have an exponential factor:
\begin{equation}
\label{eqn:ansatz}
   \mu\, I_N(A,B, \bar{A}, \bar{B})= \Psi_N\sim \sum_{\pi\in S_N} c_\pi \prod_i\,e^{a_i \bar{a}_{\pi(i)}+ b_i \bar{b}_{\pi(i)}} \;\bold{\chi}(a_i, b_i, \bar{a}_{\pi(i)},\bar{b}_{\pi(i)}).
\end{equation}
 By a symmetry argument it is also plausible that the numerator is also given by a determinant. We seek the missing factor in the numerator; we know that the Bethe ansatz is unlikely to provide a solution to our differential equation, because our effective potential appears to contain three-body interactions. Finding such a formula would amount to finding eigenfunctions of $\nabla_{A,B}$ along the lines of \cite{Berenstein:2004kk, Filev:2014qxa}, but in our case we are only interested in the ground state wavefunction in the effective potential. While a complete analysis is beyond the scope of this paper, we may still lay out a prescription for finding an analytical solution to our modified integral. We previously argued that such a solution must have $\mu$ as its denominator; following \cite{Berenstein:2004kk}, the problem can be simplified by rewriting the equation for $I_N$ and removing the effective potential in the Laplacian at the cost of adding first order derivative terms. Thus we may rewrite \eqref{eqn:ansatz} as:

\begin{equation}
 \Psi_N = \sum_{\pi\in S_N} c_\pi \prod_i\,e^{a_i \bar{a}_{\pi(i)}+ b_i \bar{b}_{\pi(i)}} \;\bold{\xi}(a_i, b_i, \bar{a}_{\pi(i)},\bar{b}_{\pi(i)})\mu = \sum_{\pi\in S_N} f_\pi\xi_\pi\mu,
 \end{equation}
where we have set $f_\pi = c_\pi \prod_i\,e^{a_i \bar{a}_{\pi(i)}+ b_i \bar{b}_{\pi(i)}}$. We see then that we have:
 \begin{equation}
 \begin{aligned}
   \lambda\Psi_N &=\left[\nabla_{a}^2 + \nabla_{b}^2\right] \Psi_N -\frac{1}{\mu}\left(\left[\nabla_{a}^2 + \nabla_{b}^2\right]\mu\right) \Psi_N\\
   &= \nabla_{a}\cdot\nabla_a\left(\sum_{\pi\in S_N} f_\pi\xi_\pi\mu\right)+\nabla_{b}\cdot\nabla_b\left(\sum_{\pi\in S_N} f_\pi\xi_\pi\mu\right) -\frac{1}{\mu}\left(\left[\nabla_{a}^2 + \nabla_{b}^2\right]\mu\right) \left(\sum_{\pi\in S_N} f_\pi\xi_\pi\mu\right)\\
   &= \nabla_{a}\cdot\left(\mu\sum_{\pi\in S_N} \nabla_a\left(f_\pi\xi_\pi\right) + \nabla_a(\mu)\sum_{\pi\in S_N}\left(f_\pi\xi_\pi\right)\right)+\nabla_{b}\cdot\left(\mu\sum_{\pi\in S_N} \nabla_b\left(f_\pi\xi_\pi\right) + \nabla_b(\mu)\sum_{\pi\in S_N}\left(f_\pi\xi_\pi\right)\right)\\   &-\left(\left[\nabla_{a}^2 + \nabla_{b}^2\right]\mu\right) \left(\sum_{\pi\in S_N} f_\pi\xi_\pi\right)\\
   \end{aligned}
   \end{equation}

Expanding, we arrive at:
\begin{equation}
\begin{aligned}
   \lambda\Psi_N&= 2\nabla_a\left(\mu\right)\cdot\left(\sum_{\pi\in S_N} \nabla_a\left(f_\pi\xi_\pi\right)\right) + \mu\sum_{\pi\in S_N} \nabla_a^2\left(f_\pi\xi_\pi\right)+\nabla_a^2(\mu)\sum_{\pi\in S_N}\left(f_\pi\xi_\pi\right)\\
   &+ 2\nabla_b\left(\mu\right)\cdot\left(\sum_{\pi\in S_N} \nabla_b\left(f_\pi\xi_\pi\right)\right) + \mu\sum_{\pi\in S_N} \nabla_b^2\left(f_\pi\xi_\pi\right)+\nabla_b^2(\mu)\sum_{\pi\in S_N}\left(f_\pi\xi_\pi\right)\\   &-\left(\left[\nabla_{a}^2 + \nabla_{b}^2\right]\mu\right) \left(\sum_{\pi\in S_N} f_\pi\xi_\pi\right)\\
   &= 2\nabla_a\left(\mu\right)\cdot\left(\sum_{\pi\in S_N} \nabla_a\left(f_\pi\xi_\pi\right)\right) + \mu\sum_{\pi\in S_N} \nabla_a^2\left(f_\pi\xi_\pi\right) + 2\nabla_b\left(\mu\right)\cdot\left(\sum_{\pi\in S_N} \nabla_b\left(f_\pi\xi_\pi\right)\right) + \mu\sum_{\pi\in S_N} \nabla_b^2\left(f_\pi\xi_\pi\right)
\end{aligned}
\end{equation}

We compute the terms containing the second derivatives and find:
\begin{equation}
\begin{aligned}
   \lambda\Psi_N
   &=2\sum_{\pi\in S_N} \left(f_\pi\nabla_a\mu\cdot\nabla_a\xi_\pi+\xi_\pi\nabla_a\mu\cdot\nabla_a f_\pi+f_\pi\nabla_b\mu\cdot\nabla_b\xi_\pi+\xi_\pi\nabla_b\mu\cdot\nabla_b f_\pi\right)\\
   &+\mu\sum_{\pi\in S_N}\left(f_\pi\nabla^2_a\xi_\pi+f_\pi\nabla^2_b\xi_\pi+\xi_\pi\nabla^2_af_\pi+\xi_\pi\nabla^2_bf_\pi+2\nabla_a\xi_\pi\cdot\nabla_a f_\pi+2\nabla_b\xi_\pi\cdot\nabla_b f_\pi\right)
\end{aligned}
\end{equation}

We know that $\mu\sum_{\pi\in S_N}\xi_\pi\nabla^2_af_\pi+\xi_\pi\nabla^2_bf_\pi = \lambda\Psi_N$. Thus we can simplify our differential equation:

\begin{equation}
\label{eqn:maindifferentialequation}
\begin{aligned}
0 &= 2\sum_{\pi\in S_N} \left(f_\pi\nabla_a\mu\cdot\nabla_a\xi_\pi+\xi_\pi\nabla_a\mu\cdot\nabla_a f_\pi+f_\pi\nabla_b\mu\cdot\nabla_b\xi_\pi+\xi_\pi\nabla_b\mu\cdot\nabla_b f_\pi\right)\\
&+\mu\sum_{\pi\in S_N}\left(f_\pi\nabla^2_a\xi_\pi+f_\pi\nabla^2_b\xi_\pi+2\nabla_a\xi_\pi\cdot\nabla_a f_\pi+2\nabla_b\xi_\pi\cdot\nabla_b f_\pi\right)
\end{aligned}
\end{equation}

We see then that we have a multivariable PDE; while there are numerical methods to approximate an analytical solution, they are all computationally laborious and ill-suited to solving PDEs containing more than two independent variables. Nevertheless, given the nature of the HCIZ integral, we note that there is another method to compute an analytical expression for the missing factor that, while tedious, is still tractable. That is to evaluate the integral explicitly and use this to determine the missing factors. We solve the integral for $U(3)$ in Appendix \ref{sec:appendixA}. We will reproduce the result directly below:
\begin{equation}
\label{eqn:su3series}
\begin{aligned}
    I_{3} &= \sum_{q=0}^\infty\sum_{\substack{f+2p+4k=q\\f,p,k\geq0}}\sum_{m,n=0}^k\sum_{\substack{j+l=n\\0\leq j\leq m\\0\leq l \leq k-m}}\sum_{g=0}^{p+k}\sum_{\substack{r+s+t+h=g\\0\leq r\leq p\\0\leq s\leq j\\0\leq t\leq m-j\\0\leq h\leq k-m}}\frac{16\pi e^{w_1}}{(k+1)(k+2)(p+4k-2m-2n+2)(q-2m-2g+2)}\\
    &\times\frac{(-1)^{-l-h}w_2^fw_3^rw_4^{p-r}w_5^sw_6^{j-s}w_7^tw_8^{m-j-t}w_9^{2(k-m)}}{2^{2(k-m)+1}f!(p-r)!r!(j-s)!s!(m-j-t)!t!l!h!(k-m-l)!(k-m-h)!}\\
\end{aligned}
\end{equation}

The expressions for $w_i$ are listed in \eqref{eqn:wexpressions}; we see immediately that the $a_i$ and $\bar{a}_j$ eigenvalues do not mix with the $b_i$ and $\bar{b}_j$ eigenvalues. We make note of the $SU(2)$ symmetry between $a_i$ and $b_i$, and $\bar{a}_j$ and $\bar{b}_j$; we note that if we remove the $b_i$ and $\bar{b}_j$, we recover the expression for the two-matrix Harish-Chandra integral evaluated over the $U(3)$ group. Given that we may remove $b_i$ and $\bar{b}_j$, replace $a_i$ and $\bar{a}_j$ with $u_i = \begin{pmatrix}a_i\\b_i\end{pmatrix}$ and $\bar{u}_j = \begin{pmatrix}\bar{a}_j\\\bar{b}_j\end{pmatrix}$, and still recover the same expressions for $w_i$, we could naively expect to recover an expression similar to the original Harish-Chandra integral \cite{itzykson1980planar}:
\begin{equation}
\label{eqn:wrongformula}
 I(u, \bar u)= \Omega \frac{\det\left(\exp(u_i \bar{u}_j)\right)}{\Delta(u)\Delta(\bar u)  }
\end{equation}
if we use $u_i$ and $\bar{u}_j$ in lieu of $a_i$, $\bar{a}_j$, $b_i$, and $\bar{b}_j$; set $\Omega$ as the normalization constant; and specify that when multiplying the Vandermondes in the denominator, one must take the dot product of $(u_i-u_j)$ and $(\bar{u}_i-\bar{u}_j)$. But upon this substitution, we find that we only recover:
\begin{equation}
 I(u, \bar u)= \Omega \frac{\det\left(\exp(u_i \bar{u}_j)\right)}{\mu}
\end{equation}
We are missing the factors of $\chi(a_i,b_i,\bar{a}_j,\bar{b}_j)$ in the numerator. We can solve for the missing factors by examining the case of $U(3)$; given that the effective potential is the same general form for $N > 2$, we may extrapolate the missing factor for a general $U(N)$ formula from the $U(3)$ results. 

In the case of $U(3)$, we can expand the original HCIZ integral:
\begin{equation}
\label{eqn:originalHCIZ}
 I(\Lambda, \bar \Lambda)= \int d U\exp\left(\Tr\left(  U ^{-1}\Lambda U \bar \Lambda\right)\right) = \Omega \frac{\det\left(\exp(\lambda_i \bar\lambda _j)\right)}{\Delta(\Lambda)\Delta(\bar \Lambda)  }
\end{equation}
until we arrive at a series that takes the form of \eqref{eqn:su3formula} with modified $w_i$ to reflect the omission of the $b_i$ and $\bar{b}_j$. We examine the effects of replacing $a_i$ and $a_j$ with $u_i$ and $u_j$ at each step; we then compare the results to an expansion of \eqref{eqn:wrongformula}. The additional terms that replacing $a_i$ and $a_j$ with $u_i$ and $u_j$ yields must sum up to the missing factors. We briefly sketch out the start of such an expansion. We note that for $SU(3)$, we have:
\begin{equation}
\begin{aligned}
\det(\exp(a_i\bar{a}_j + b_i\bar{b}_j)) &= -e^{a_3\bar a_1 + a_2\bar a_2 + a_1\bar a_3 + b_3\bar b_1 + b_2\bar b_2 + b_1\bar b_3}+e^{a_2\bar a_1 + a_3\bar a_2 + a_1\bar a_3 + b_2\bar b_1 + b_3\bar b_2 + b_1\bar b_3}\\
&+e^{a_3\bar a_1 + a_1\bar a_2 + a_2\bar a_3 + b_3\bar b_1 + b_1\bar b_2 + b_2\bar b_3}-e^{a_1\bar a_1 + a_3\bar a_2 + a_2\bar a_3 + b_1\bar b_1 + b_3\bar b_2 + b_2\bar b_3}\\
&-e^{a_2 \bar a_1 + a_1 \bar a_2 + a_3\bar a_3 + b_2 \bar b_1 + b_1 \bar b_2 + b_3 \bar b_3} + e^{a_1 \bar a_1 + a_2\bar a_2 + a_3\bar a_3 + b_1\bar b_1 + b_2\bar b_2 + b_3\bar b_3}\\
&= -e^{s_1} + e^{s_2} +  e^{s_3} - e^{s_4} - e^{s_5} + e^{s_6}
\end{aligned}
\end{equation}
We have set:
\begin{equation}
\begin{aligned}
s_1 &= a_3\bar a_1 + a_2\bar a_2 + a_1\bar a_3 + b_3\bar b_1 + b_2\bar b_2 + b_1\bar b_3\\
s_2 &= a_2\bar a_1 + a_3\bar a_2 + a_1\bar a_3 + b_2\bar b_1 + b_3\bar b_2 + b_1\bar b_3\\
s_3 &= a_3\bar a_1 + a_1\bar a_2 + a_2\bar a_3 + b_3\bar b_1 + b_1\bar b_2 + b_2\bar b_3\\
s_4 &= a_1\bar a_1 + a_3\bar a_2 + a_2\bar a_3 + b_1\bar b_1 + b_3\bar b_2 + b_2\bar b_3\\
s_5 &= a_2 \bar a_1 + a_1 \bar a_2 + a_3\bar a_3 + b_2 \bar b_1 + b_1 \bar b_2 + b_3 \bar b_3\\
s_6 &= a_1 \bar a_1 + a_2\bar a_2 + a_3\bar a_3 + b_1\bar b_1 + b_2\bar b_2 + b_3\bar b_3
\end{aligned}
\end{equation}
If we expand each term as a Taylor series, we may rewrite our determinant as:
\begin{equation}
\begin{aligned}
\det(\exp(a_i\bar{a}_j + b_i\bar{b}_j)) &= \sum_m \frac{1}{m!}(s_2 - s_1)\sum_{n=0}^{m-1}s_2^ns_1^{m-n-1}\\
&- \sum_m \frac{1}{m!}(s_4 - s_3)\sum_{n=0}^{m-1}s_4^ns_3^{m-n-1}\\
&+ \sum_m \frac{1}{m!}(s_6 - s_5)\sum_{n=0}^{m-1}s_6^ns_5^{m-n-1}
\end{aligned}
\end{equation}
We note that:
\begin{equation}
\label{eqn:bfactors}
\begin{aligned}
s_2 - s_1 &= (a_2-a_3)(\bar a_1 - \bar a_2) + (b_2-b_3)(\bar b_1 - \bar b_2)\\
s_4 - s_3 &= (a_1-a_3)(\bar a_1 - \bar a_2) + (b_1-b_3)(\bar b_1 - \bar b_2)\\
s_6 - s_5 &= (a_1 - a_2)(\bar a_1 - \bar a_2) + (b_1 - b_2)(\bar b_1 - \bar b_2)
\end{aligned}
\end{equation}

We see immediately that if we remove the $b_i$ and $b_j$, the factors listed above cancel out factors in the Vandermonde determinants of the original HCIZ integral; however, once we add in $b_i$ and $b_j$, some of the factors in \eqref{eqn:bfactors} no longer cancel factors in $\mu$. This suggests that there are missing saddle points and that the missing factors should add the terms needed to restore the overall factor of $\mu$ in the numerator. We note that our results should generalize to an arbitrary number of matrices; we would simply modify $\mu$ to account for the additional matrices and add the relevant derivative terms to \eqref{eqn:maindifferentialequation}, as well as modify $u_i$ and $\bar{u}_j$ to account for the new eigenvalues. 

We leave off here and save this computation for future works. 
%%%%%%%%%%%%%%%%%%%%%%%%%%%%%%%%%%%%%%%%%%%%%%%%%%%%%%%

\section{Connection with Restricted Schur Polynomials and Collective Coordinates}
\label{sec:third}

A natural question to ask is: what sort of basis of operators do the coherent states \eqref{eqn:twomatrixstate} actually generate? This is quite non-trivial, since there are in principle many different ways of orthogonalizing the two point function of $\frac{1}{4}$-BPS operators at finite $N$. As a concrete example, we can take the simplest multi-matrix coherent state we obtain from choosing the coherent state parameters to be rank one projectors for arbitrary $N$. These states will describe semi-classical configurations of single quarter BPS giant gravitons. Similar generating functions were introduced in \cite{Chen:2019gsb, Lin:2022gbu};  we will clarify the relationship between them and the coherent states studied here. This should help in generalizing to higher rank cases corresponding to bound states of AdS giants. The idea is to consider the following state:
\begin{equation}
    \ket{\lambda_x, \lambda_y}=\int_{\mathbb{CP}^{N-1}} d\varphi^\dagger d\varphi\;e^{\lambda_x \varphi^\dagger X \varphi+ \lambda_y \varphi^\dagger Y \varphi}\ket{0}.
\end{equation}
To evaluate this, we need a formula for the moments of $\varphi^\dagger_i \varphi_j$ with respect to the flat Haar measure on $\mathbb{CP}^{N-1}$. The measure can be rewritten as follows:
\begin{equation}
   \int_{\mathbb{CP}^{N-1}} d\varphi^\dagger d\varphi= \int_{i \mathbb{R}} ds\; \int d\bar{\phi} d \phi\; e^{-s\left(\bar{\phi}\phi-1 \right)} = \int_{\mathcal{C}} e^{s}\; \int \ d\bar{\phi} d \phi\; e^{-s\bar{\phi}\phi}.
\end{equation}
In other words, we can trade the integral over projective space for a regular Gaussian integral at the cost of introducing an additional contour integral over an auxiliary parameter $s$. Then the moments have a simple expression in terms of the projection operators $P_{(k)}$ \cite{Chen:2019gsb}:
\begin{equation}
     \int_{\mathbb{CP}^{N-1}} d\varphi^\dagger \prod_{l=1}^k\left(\varphi^\dagger\right)^{i_l} \varphi_{j_l}= \oint  \frac{ds\,e^s}{s^{N+k}}\; k! \left( P_{(k)}\right)^I_J= \frac{k!}{(N+k-1)!} \left( P_{(k)}\right)^I_J.
\end{equation}
Borrowing the results of \cite{Chen:2019gsb}, we may rewrite the coherent state as a sum of the so-called restricted Schur polynomial operators $\chi_{(k_1+k_2), (k_1)\, (k_2)}(X,Y)$:
\begin{equation}
  \ket{\lambda_x, \lambda_y}= \sum_{k_1=0}^\infty \sum_{k_2=0}^\infty\; \frac{\lambda_x^{k_1} \lambda_2^{k_2}}{(N+k_1+k_2-1)!}\chi_{(k_1+k_2), (k_1)\, (k_2)}(X,Y)\ket{0}.
\end{equation}
Now we would like to understand the analogue of this formula in the general case.

First, we need to recall the definition of the restricted Schur polynomials:
\begin{equation}
    \chi_{R,(r,s)\, \alpha \beta}(X,Y)= \Tr[P_{R,(r,s)\, \alpha \beta}\,X^n \otimes Y^m].
\end{equation}
Here, $R$ is a Young diagram associated with an irreducible representation of $S_{n+m}$; the labels $(r,s)$ correspond to an irreducible representation of $S_n\times S_m$ contained in $R$. The object $P_{R,(r,s)\, \alpha \beta}$ can be understood as follows: starting with $S_m\times S_n\subset S_{m+n}$, we can find representations $r\times s$ sitting within $R$. Generically, the representation $r\times s$ can appear more than once inside of $R$, so one needs to keep track of how one embeds $r\times s$ into $R$. If the multiplicity of $(r,s)$ is $n_{(r,s)}$ and its dimension $d_{(r,s)}$, then a generic element of $S_{n+m}$ will be block diagonalized into  $\left(n_{(r,s)} d_{(r,s)}\right)\times \left(n_{(r,s)} d_{(r,s)}\right)$ blocks. The matrix indices $\alpha, \beta$ keep track of this information, where $\alpha$ and $\beta$ range from $1$ to $n_{(r,s)}$. The $P_{R,(r,s)\, \alpha \beta}$ are then intertwining operators between each of these blocks. More formally, we can label each of the embeddings of $r\times s$ by an index $\gamma$ and consider the space $R_\gamma \subset R$. The restricted Schur polynomial is then given by:
\begin{equation}\label{eqn:restrictedSchur}
     \chi_{R,R_{\gamma}}(X,Y) = \frac{1}{m! n!}\sum_{\sigma \in S_{n+m}} \Tr_{R_{\gamma}}[\Gamma_R(\sigma)] \Tr[\sigma X^n\otimes Y^m],
\end{equation}
where $\Gamma_R(\sigma)$ is the matrix representing $\sigma$ \cite{deMelloKoch:2007rqf}. The most complicated part of the restricted Schur polynomials is the evaluation of $\Tr_{R_{\gamma}}[\Gamma_R(\sigma)]$, which involves building $R_\gamma$ explicitly.

By expanding the exponential and evaluating the unitary integrals, we obtain:
\begin{equation}\label{eqn: twomatrixexpansion}
\begin{aligned}
&\frac{1}{\text{Vol}\left[U(N)\right]} \int dU\; \exp\left(U X U^\dagger \Lambda_X +UY U^\dagger \Lambda_Y\right)=\\ & \sum_{n,m}\frac{1}{m! n!} \sum_{\sigma, \tau \in S_{n+m}} \Tr[\sigma \Lambda_X^n\otimes \Lambda_Y^m]\Tr[\tau X^n\otimes Y^m] \, \boldsymbol{Wg}(\sigma \tau^{-1}, N) ,
\end{aligned}
\end{equation}
where $\boldsymbol{Wg}(\sigma, N)$ is the Weingarten function. Explicit combinatorial formulas for Weingarten functions are well known from the work of Collins (see \cite{collins2021weingarten} for an elementary introduction); before delving into specific details, we should contrast this with the situation where one of the $\Lambda_{X,Y}$ is zero. In this case, the resulting sum can be recast as a diagonal sum of the products of unitary characters; right now, we have a complicated sum of traces. For a moment, let us consider the situation for a single matrix. The resulting sum is:
\begin{equation}
\begin{aligned}
&\frac{1}{\text{Vol}\left[U(N)\right]} \int dU\; \exp\left(U X U^\dagger \Lambda_X\right)\\ &= \sum_{n=0}^\infty\frac{1}{ n!} \sum_{\sigma, \tau \in S_{n}} \Tr[\sigma \Lambda_X^n]\Tr[\tau^{-1} X^n] \, \boldsymbol{Wg}(\sigma \tau^{-1}, N) \\
&= \sum_{n=0}^\infty\frac{1}{ n!} \sum_{\sigma, \tau \in S_{n}} \Tr[\sigma \Lambda_X^n]\Tr[\tau^{-1} X^n] \, \sum_{\lambda\vdash n} \frac{1}{n!\,f_\lambda} \chi^\lambda(\tau^{-1}\sigma) \chi^\lambda(1)\\
&= \sum_{n=0}^\infty \sum_{\lambda\vdash n} \frac{1}{f_\lambda} s_\lambda(X) s_\lambda(\Lambda_X).
\end{aligned}
\end{equation}

The last line is obtained from the character expansion of the integral, which was computed in \cite{Berenstein:2022srd}. Then for two matrices, we have:
\begin{equation}
\begin{aligned}
&\frac{1}{\text{Vol}\left[U(N)\right]} \int dU\; \exp\left(U X U^\dagger \Lambda_X +UY U^\dagger \Lambda_Y\right)\\ &= \sum_{n,m}\frac{1}{m! n!(n+m)!} \sum_{\lambda\vdash n+m}\frac{1}{f^\lambda} \sum_{\sigma, \tau \in S_{n+m}} \chi^\lambda(\sigma) \chi^\lambda(\tau)\Tr[\sigma \Lambda_X^n\otimes \Lambda_Y^m]\Tr[\tau X^n\otimes Y^m] .
\end{aligned}
\end{equation}

Clearly this has a similar structure to the definition of the restricted Schur polynomials \eqref{eqn:restrictedSchur}, but the restricted characters have been replaced with ordinary symmetric group characters instead. We can always formally rewrite each of the terms in the series as a sum over restricted characters by decomposing the trace over $R$ into a sum of traces over each of the $R_\gamma$ : 
\begin{equation}
    \chi^R(\sigma)= \Tr_R[\Gamma_R(\sigma)]= \sum_{\alpha} \sum_{(r,s)_{\alpha \alpha}\subset R} \Tr_{(r,s)_{\alpha \alpha}}[\Gamma_R(\sigma)].
\end{equation}
This allows us to rewrite the integral as a sum of restricted Schur polynomial operators.
However, this sum does not capture every restricted character; the problem comes from elements $\sigma \in S_{n+m}$ that are not elements of $S_n\times S_m$. Generically, the representation $R$ of $S_n$ will be expressed as a sum of the irreducible representations $R_{\alpha}$ of $S_n\times S_m$ with multiplicities $n_\alpha$. This means that $\sigma$ is not fully block diagonal on the space $n_{\alpha} R_{\alpha}$, and indeed, restricted traces on each of these blocks lead to different orthogonal states in the free theory. This is expected, since the coherent state only generates operators that have vanishing one-loop anomalous dimension, and the dilatation operator acts non-trivially on generic restricted Schur polynomial operators. This means that the operators obtained from diagonal traces over $n_{\alpha} R_{\alpha}$ have vanishing one-loop anomalous dimension, while off-diagonal operators should be associated with excited states with open strings. We expect that operators that are approximate eigenstates of the dilatation operator in the large $N$ limit take the form of open string modifications of these coherent states, and are likely more closely related to the Gauss graph operators as in \cite{deMelloKoch:2012ck}. 

An interesting direction to take would be to construct a generating function for all restricted Schur polynomials. Clearly, something as simple as \eqref{eqn:twomatrixstate} cannot work. This can be traced back to the fact that the sum over $S_{n+m}$ has many redundancies owing to the fact that we can conjugate by an element of $S_n\times S_m$ while leaving the traces fixed. This is the statement that we can permute the $n$ $X$'s and $m$ $Y$'s among themselves while simultaneously permuting the $\Lambda_{X,Y}$'s. As explained in \cite{Bhattacharyya_2008}, there is an equivalence relation between elements of $S_{n+m}$ in such a way that
\begin{equation}
    \sigma \sim \tau \, \Leftrightarrow \Tr[\sigma A^n \otimes B^m]=\Tr[\tau A^n \otimes B^m],
\end{equation}
which happens exactly when $\sigma$ can be conjugated into $\tau$ by an element of $S_n\times S_m$. In other words, the construction of restricted Schur polynomials is equivalent to constructing generalized class functions on restricted conjugacy classes, which means that the coherent state generating function \eqref{eqn:twomatrixstate} cannot differentiate between different restricted Schur polynomials by itself for the simple reason that the Weingarten function is a class function. If we want to replace the characters in \eqref{eqn:twomatrixstate} with restricted characters, we must either change the domain of integration or integrate against an appropriate measure factor that is sensitive to this information. This is equivalent to finding an analytic formula for restricted characters, which may be recast as a Schr\"odinger problem over the space of commuting matrices \cite{Berenstein:2005aa, Filev:2014qxa}. The point is that the norm of the quarter-BPS coherent state is related to the heat kernel over the space of commuting matrices, or equivalently to the Green's function of the Schr\"odinger equation. 
In practice, the coherent states still form an overcomplete basis of operators that can be used for computations, even if we do not currently know how to project into a particular primary state; the leading contribution at large $N$ will come from the saddle point approximation. 

In Appendix \ref{sec:appendixB}, we checked a few low order terms by explicit calculation and found that the quarter-BPS coherent state generating function is given by a sum of product of restricted Schur polynomials much like the half-BPS coherent state. As we explained, this should hold for all the terms in the series, but checking higher order terms is difficult. This can be taken as further evidence that the coherent states span all possible BPS states and makes manifest many of the ideas in \cite{Berenstein:2005aa}, since free field theory correlators can be encoded in integrals of polynomial functions of the collective coordinates. In the saddle point approximation, the integration measure is given by 
\begin{equation}\label{intmeasure}
   \int d\vec{\Lambda} \cdot d \vec{\bar{\Lambda}}\, \bra{\bar{\Lambda}_X, \bar{\Lambda}_Y}\ket{\Lambda_X, \Lambda_Y} \simeq \int \prod_i\,d\lambda^x_i d\bar{\lambda}^x_id\lambda^y_i d\bar{\lambda}^y_i\; \prod_{i<j} \left(\Vec{\lambda}_i-\Vec{\lambda}_j\right) \left(\Vec{\bar{\lambda}}_i-\Vec{\bar{\lambda}}_j\right) \;e^{\vec{\lambda
    }_i\cdot\vec{\bar{\lambda}
    }_i}.
\end{equation}
After an appropriate choice of contour for which $\lambda$ and $\bar{\lambda}$ are canonically conjugate variables, this reproduces the strong coupling ansatz in \cite{Berenstein:2005aa}. It should also be clear that this measure describes the ground state wavefunction. Although our analysis is strictly on the weak coupling regime, we note that the dilatation operator in the $SU(2)$ sector can only act by permutations; this sector is closed, so it is very plausible that the quarter-BPS states are not renormalized. Even at weak coupling, the vacuum structure becomes quite non-trivial, modifying the Coulomb branch analysis for small collective coordinates, since the eigenvalues behave as strongly coupled bosons at low energies. This modifies the topology of the moduli space of vacua near the center of mass of the eigenvalues, even for half-BPS configurations. Since this sector preserves as much supersymmetry as $\mathcal{N}=2$ SYM, it is quite plausible that the $g_{YM}$ corrections are under control for sufficiently small modifications of BPS operators. At strong coupling, the expectation is that such states describe rotating strings propagating on bubbling geometries; the energy of these strings will follow the dispersion relation of a centrally extended BPS state
\begin{equation}
    \Delta-J =\sqrt{Q^2+ |M|^2},
\end{equation}
with the central charge $M$ being related to the length of the string in the bubbling geometry times the string tension. This is purely a kinematic effect; all of the dynamics should be encoded in the central charge $M$. Near the core of the geometry, the naive Coulomb branch analysis certainly breaks down, but S-duality considerations suggest that the string tension is not renormalized \cite{Berenstein:2009qd}, so the corrections to curvature on the moduli space should come from finite $N$ effects. This should be captured by non-planar corrections involving the exchange of gravitons between the background and a probe string. Such geometries can be engineered by integrating against the wavefunctions that break the $SO(6)_R$ symmetry of the vacuum. It is not hard to come up with such wavefunctions (for instance a non-symmetric Gaussian perturbation), and there is a schematic mapping between the eigenvalue distribution at large $N$ and bubbling geometry \cite{Chen:2007du}. The relation \eqref{intmeasure} should be corrected with additional $1/N$ effects, since the naive saddle point approximation fails to give the exact overlap, but these effects should only be relevant when we try to probe eigenvalues are placed in non-generic configurations. These should be thought of as microstate configurations for coarse-grained eigenvalue droplet configurations associated to superstar geometries. For $\frac{1}{8}$ BPS states, the analysis is more subtle; we have to take into account the effects from fermions since we would be working in a $SU(2|3)$ subsector. One should be able to ignore the effect of the fermions for large enough semiclassical operators. This makes this class of coherent states ideal for studying near-BPS limits around large operators without having to deal with the mixing of multi-trace structures. 

\section{Discussion}
\label{sec:conclusion}

In this paper, we studied multi-matrix coherent states for bosonic matrices that generate $\frac{1}{4}$ and $\frac{1}{8}$ BPS states in $\mathcal{N}=4$ SYM. We showed that the norm of these coherent states admits a fixed point formula generalizing the Harish-Chandra-Itzykson-Zuber formula for gauge group $U(2)$, and provided evidence of an expansion in terms of restricted Schur polynomials for $U(N)$. This gives in principle a way of generating expressions for BPS states for any value $N$ in $\mathcal{N}=4$ SYM. One technical obstacle we face is that our construction does not give an alternative construction of the so-called restricted Schur polynomial operators \cite{deMelloKoch:2007rqf}. This is related to the expectation that there is a hidden symmetry under which different operators are charged. One idea is that determining the Casimir charges should be enough to differentiate between different operators, but this problem is quite non-trivial even in the $\frac{1}{2}$ BPS sector \cite{Kemp:2023kma}. It is also unclear how to implement this idea efficiently at large $N$ since the number of Casimirs needed to distinguish between different operators grows with the complexity of the operators. Despite this obstacle, our results are important for computing correlators of $\frac{1}{4}$ and $\frac{1}{8}$ BPS operators dual to bound states of giant gravitons \cite{Mikhailov:2000ya} and generic bubbling geometries \cite{Chen:2007du}. Understanding the precise map between the overcomplete 'eigenvalue basis' of coherent states and specific orthogonal bases of operators remains an important problem. We conclude with a few more immediate directions for future work.
 
\subsection*{$\frac{1}{16}$ BPS States and Black Hole Microstate Operators}

One of the more interesting generalizations would be to the case of $\frac{1}{16}$ BPS operators. By now, there is ample evidence that there exists a class of $\frac{1}{16}$ BPS operators describing the microstates of supersymmetric black holes in $AdS_5 \times S^5$ \cite{Benini:2018ywd, Cabo-Bizet:2018ehj, Chang:2022mjp, Choi:2023znd}. Recently, there have been some studies of these types of states for small values of $N$ \cite{Budzik:2023vtr, Chang:2023zqk}; see \cite{Budzik:2023xbr} for a more general discussion. our results imply that one should be cautious in extrapolating results about operators for the $U(2)$ theory, since multi-matrix states appear to be qualitatively different for other values of $N$. We expect that most of the interesting qualities of such operators are missing from the $U(2)$ and $SU(2)$ theory. It would be nice to develop more systematic techniques to build these types of operators. In principle, there are no obstructions to generalizing our techniques to this setup, with the working assumption that finding states with vanishing one-loop anomalous dimension is enough \cite{Grant:2008sk}. The idea would essentially be to build a superfield coherent state \cite{chang20131}:
\begin{equation}\label{eqn: superfield coherent state}
    \int dU \; \exp\left\{\int d^3 \theta \int dz \Tr\left[U\Psi U^\dagger \Phi \right] \right\} \ket{0},
\end{equation}
where $\Psi(z, \theta)$ is the $\mathbb{C}^{2|3}$ superfield discussed in \cite{chang20131, Chang:2022mjp}, and $\Phi$ is an auxiliary superfield of coherent state parameters. The combined effect of the exponentiation and integration over the unitary matrices is to generate all possible gauge invariant tensor contractions. One should expect that the operators generated by this generating function are generalizations of the $SU(2|3)$ restricted Schur polynomials constructed in \cite{deMelloKoch:2012sie}. Generically, the terms in the expansion of \eqref{eqn: superfield coherent state} will not be of multi-graviton form, so they are natural candidates for microstates of supersymmetric black holes. In practice, the main disadvantage of an expression like \eqref{eqn: superfield coherent state} is that it might not be practically useful, in the sense that the expansion necesarily involves an infinite number of matrix fields associated with covariant derivatives acting on the fields. One way of avoiding this difficulty is to use generating functions such as the ones studied in \cite{Lin:2022wdr}. Alternatively, one can view the auxiliary superfield $\Phi$ as a full-fledged dynamical collective coordinate. One would then hope that integrating out the SYM fields leads to an effective matrix quantum mechanics describing (near)-BPS black hole microstates, with the lightcone coordinate $z$ acting as a time variable.

\subsection*{Three Point Correlators, Bubbling Geometries, and Twisted Holography}

Although eventually we would like to study black holes, it is important to build intuition from simpler examples. One class of such examples is the BPS bubbling geometries \cite{Chen:2007du} generalizing LLM geometries \cite{Lin:2004nb}. Although the droplet description of such states in supergravity is compelling, a precise mapping between the weak coupling BPS states is not fully developed\footnote{For instance, it is unclear whether the solutions found in \cite{Chen:2007du} exhaust the set of all $\frac{1}{4}$ and $\frac{1}{8}$ BPS states.}. The coherent states \eqref{eqn: twomatrixexpansion} have a more natural connection to such geometries\cite{Holguin:2023orq}. A worthwhile exercise would be to study correlators of single trace chiral primaries in the background of heavy coherent states corresponding to both giant gravitons or bubbling geometries; see \cite{suzuki2020three} for some finite $N$ results. The holographic renomalization techniques of \cite{Skenderis:2007yb} are also applicable in these cases, but it would be interesting to develop more efficient computational techniques in supergravity along the lines of \cite{abajian2023holography}. A good toy model for this would be to study these types of questions in Twisted Holography \cite{Costello:2018zrm}. In that set-up, the eigenvalue droplets should be related to bubbling on a 2 dimensional complex base with holomorphic coordinates $X,Y$, with the vacuum configuration being a droplet with the topology of $S^3$; this is the deformed conifold description of $SL(2, \mathbb{C})$. More generic eigenvalue configurations should lead to other non-compact Calabi-Yau threefolds such as in \cite{Gomis:2006mv, Halmagyi:2007rw}. The expectation is that the geometry on both sides of the duality is encoded by a spectral curve  \cite{Budzik:2021fyh} on which stacks of branes are supported, which should appear as the spectral curve of the corresponding matrix model. Understanding this could help in clarifying the dictionary between collective coordinates and bubbling geometries in the $AdS_5$ case.

\subsection*{Bound States of Giants and Branes at Angles}

One reasonable goal would be to understand coherent states associated with pairs of eigenvalues, which are built from simpler integrals over the Grassmanian $Gr(2, N)$. The main difficulty of such a character expansion already appears in this simpler case:
\begin{equation}
  \ket{\lambda_x, \lambda_y}=  \int_{Gr(2, N)} dV^\dagger dV \; e^{\Tr[VXV^\dagger \lambda_x +VYV^\dagger \lambda_y]},
\end{equation}
where the $\lambda_\alpha$ are not $2\times 2$ diagonal matrices. The norm of this state has a rather explicit integral form over $2\times 2$ matrices over a compact domain; thus an explicit evaluation should be feasible. We expect that this state has a non-trivial expansion in terms of restricted Schur polynomials \cite{Chen:2019gsb}. Since the domain of integration is simpler than that of the general case, the integral might lead itself to a saddle point analysis. One might be able to explicitly compute Lefshetz thimbles for this case and determine whether localization fails or not, or whether complex saddle points are needed. It would be interesting to construct multi-matrix analogues of the generating functions found in \cite{Carlson:2022dot} for determinant operators. This would help in constructing the precise operators dual to intersecting giants \cite{Holguin:2021qes}, and particularly in understanding their integrable boundary states \cite{Piroli:2017sei}.

\acknowledgments

We would like to thank D. Berenstein for helpful discussions. A.H. would like to thank the organizers of the Physics Summer workshop at the Simons Center for Geometry and Physics for their hospitality. SW's research was supported in part by the Department of Energy under grant DE-SC0019139.

\appendix
\section{The Four-Matrix Model in $U(3)$}
\label{sec:appendixA}

We now consider the following integral:
\begin{equation}
\label{U3exponential}
    I = \int dU(3)\exp(\Tr(\bar{A}UA U^{\dagger}) + \Tr(\bar{B}UB U^{\dagger})),
\end{equation}

for $A = \begin{pmatrix}a_1 \\ a_2\\ a_3\end{pmatrix}$, $\bar{A} = \begin{pmatrix}\bar{a}_1 & \bar{a}_2 & \bar{a}_3\end{pmatrix}$, $B = \begin{pmatrix}b_1 \\ b_2 \\ b_3\end{pmatrix}$, and $\bar{B} = \begin{pmatrix}\bar{b}_1 & \bar{b}_2 & \bar{b}_3\end{pmatrix}$.

We know that we can parameterize our $U(3)$ matrix as:
\begin{equation}
U = e^{i\lambda_3\alpha}e^{i\lambda_2\beta}e^{i\lambda_3\sigma}e^{i\lambda_5\theta}e^{i\lambda_3 a}e^{i\lambda_2 b}e^{i\lambda_3 c}e^{i\lambda_8\phi},
\end{equation}
where $\lambda_i$ denotes the $i$th generators of $U(3)$. We list the relevant $SU(3)$ generators below \cite{1997physics...8015B}:
\begin{equation}
    \lambda_2 = \begin{pmatrix}0 & -i & 0\\i & 0 & 0\\ 0 & 0 & 0\end{pmatrix}, \lambda_3 = \begin{pmatrix}1 & 0 & 0\\0 & -1 & 0\\ 0 & 0 & 0\end{pmatrix}, \lambda_5 = \begin{pmatrix}0 & 0 & -i\\0 & 0 & 0\\ i & 0 & 0\end{pmatrix}, \lambda_8 = \frac{1}{\sqrt{3}}\begin{pmatrix}1 & 0 & 0\\0 & 1 & 0\\ 0 & 0 & -2\end{pmatrix}
\end{equation}

Because $U(3) = SU(3)\times U(1)$, we multiply $U$ by an additional phase $e^{i\psi}$. This yields the parameterization of $U(3)$ that we will use to compute the argument of the exponential in Eq. \eqref{U3exponential}. Simplifying and expanding, we arrive at:
\begin{equation}
\begin{aligned}
    \Tr\left(\bar{A}UAU^\dagger+\bar{B}UBU^\dagger\right) &= (a_3\bar{a_3} + b_3\bar{b_3}) \cos^2\theta + (a_2\bar{a_2} + b_2\bar{b_2})\cos^2\beta\cos^2 b\\
    &+ (a_1\bar{a_1} + b_1\bar{b_1})\cos^2\beta\cos^2\theta\cos^2 b + (\bar{a_1}a_2 + \bar{b_1}b_2)\cos^2 b\sin^2\beta\\ 
    &+ (a_1\bar{a_2} + b_1\bar{b_2})\cos^2\theta\cos^2 b\sin^2\beta + (\bar{a_1}a_3 + \bar{b_1}b_3)\cos^2\beta\sin^2\theta\\ 
    &+ (a_1\bar{a_3} + b_1\bar{b_3})\cos^2 b\sin^2\theta + (\bar{a_2}a_3 + \bar{b_2}b_3)\sin^2\beta\sin^2\theta\\
    &+ (a_1\bar{a_2} + b_1\bar{b_2})\cos^2\beta\sin^2 b + (\bar{a_1}a_2 + \bar{b_1}b_2)\cos^2\beta \cos^2\theta\sin^2 b\\ 
    &+ (a_1\bar{a_1} + b_1\bar{b_1})\sin^2\beta\sin^2 b + (a_2\bar{a_2} + b_2\bar{b_2})\cos^2\theta\sin^2\beta\sin^2 b\\
    &+ (a_2\bar{a_3} + b_2\bar{b_3})\sin^2\theta\sin^2 b\\
    &-\frac{1}{4}(a_1\bar{a_1} - \bar{a_1}a_2 - a_1\bar{a_2} + a_2\bar{a_2} + b_1\bar{b_1} - \bar{b_1}b_2 - b_1\bar{b_2} + b_2\bar{b_2})\\
    &\times e^{-2i(\sigma + a)}\cos\theta\sin2\beta\sin2b\\
    &-\frac{1}{4}(a_1\bar{a_1} - \bar{a_1}a_2 - a_1\bar{a_2} + a_2\bar{a_2} + b_1\bar{b_1} - \bar{b_1}b_2 - b_1\bar{b_2} + b_2\bar{b_2})\\
    &\times e^{2i(\sigma + a)}\cos\theta\sin2\beta\sin2b
\end{aligned}
\end{equation}

We plug this into the integral. We may use the method outlined in \cite{1997physics...8015B} to compute the Haar measure; we arrive at:
\begin{equation}
dU = \sin2\beta\sin2\theta\sin2b\sin^2\theta d\alpha d\beta d\sigma d\theta d a d b d c d\phi d\psi
\end{equation}

We cite the angle limits from \cite{1997physics...8015B}:
\begin{equation}
    \begin{aligned}
    0 &\leq \alpha, \sigma, a, c, \psi < \pi\\
    0 &\leq \beta, b, \theta < \frac{\pi}{2}\\
    0 &\leq \phi < 2\pi\\
    \end{aligned}
\end{equation}

We seek to integrate over $\sigma$ and $a$ first. We observe that:
\begin{equation}
    -e^{2i(\sigma + a)} - e^{-2i(\sigma + a)} = -2\cos(\sigma +a)
\end{equation}

Returning to our integral, we note that rather than integrate over $\sigma$, we may perform a change of variables and integrate over $\sigma + a$ and change our integration limits to $\left[0, 2\pi\right]$. We would integrate over $a$ twice, but the integral over $a$ is trivial; as long as we note that the integral over the angular variable $a$ is normalized to $1$, we may proceed with integrating. Then the relevant integral is:
\begin{equation}
\begin{aligned}
I_\sigma &= \int_0^{2\pi}\exp\left(-\frac{1}{2}(a_1\bar{a_1} - \bar{a_1}a_2 - a_1\bar{a_2} + a_2\bar{a_2} + b_1\bar{b_1} - \bar{b_1}b_2 - b_1\bar{b_2} + b_2\bar{b_2})\sin2\beta\sin2b\cos\theta\sin(\sigma+a)\right)d\sigma\\
&= 2\pi I_0\left(\frac{1}{2}(a_1\bar{a_1} - \bar{a_1}a_2 - a_1\bar{a_2} + a_2\bar{a_2} + b_1\bar{b_1} - \bar{b_1}b_2 - b_1\bar{b_2} + b_2\bar{b_2})\sin2\beta\sin2b\cos\theta\right)
\end{aligned}
\end{equation}
where $I_0$ is the modified Bessel function of the first kind of order zero. There are no factors of $a$ left in the integrand; as long as $\int_0^{\pi}da$ is normalized by a factor of $\pi$, we may consider the previous integral to have simultaneously integrated over both $\sigma$ and $a$.  

We now seek to integrate over $\theta$. First, we collect the relevant terms and group the coefficients for simplicity's sake. We have:
\begin{equation}
\begin{aligned}
    v_1 &= \sin2\beta\sin2b\\
    v_2 &= (a_2\bar{a_2} + b_2\bar{b_2})\cos^2\beta\cos^2 b + (\bar{a_1}a_2 + \bar{b_1}b_2)\cos^2 b\sin^2\beta\\ 
    &+ (a_1\bar{a_2} + b_1\bar{b_2})\cos^2\beta\sin^2 b + (a_1\bar{a_1} + b_1\bar{b_1})\sin^2\beta\sin^2 b\\
    v_3 &= (a_3\bar{a_3} + b_3\bar{b_3}) + (a_1\bar{a_1} + b_1\bar{b_1})\cos^2\beta\cos^2 b + (a_1\bar{a_2} + b_1\bar{b_2})\cos^2 b\sin^2\beta\\
    &+ (\bar{a_1}a_2 + \bar{b_1}b_2)\cos^2\beta\sin^2 b + (a_2\bar{a_2} + b_2\bar{b_2})\sin^2\beta\sin^2 b\\
    v_4 &= (\bar{a_1}a_3 + \bar{b_1}b_3)\cos^2\beta + (a_1\bar{a_3} + b_1\bar{b_3})\cos^2 b\\
    &+ (\bar{a_2}a_3 + \bar{b_2}b_3)\sin^2\beta + (a_2\bar{a_3} + b_2\bar{b_3})\sin^2 b\\
    v_5 &= \frac{1}{2}(a_1\bar{a_1} - \bar{a_1}a_2 - a_1\bar{a_2} + a_2\bar{a_2} + b_1\bar{b_1} - \bar{b_1}b_2 - b_1\bar{b_2} + b_2\bar{b_2})\sin2\beta\sin2b
\end{aligned}
\end{equation}

Our integral over $\theta$ thus becomes:
\begin{equation}
I_\theta = 2\pi\int_0^{\frac{\pi}{2}}v_1e^{v_2+v_3\cos^2\theta+v_4\sin^2\theta}I_0(v_5\cos\theta)\sin2\theta\sin^2\theta d\theta,
\end{equation}
where $I_\theta$ is the integral over $\theta$ in $I$ from eq.~\eqref{U3exponential} and $v_i$ denote the grouped coefficients. We now observe that we may rewrite $\sin2\theta$ as $2\sin\theta\cos\theta$. Then our integral becomes:

\begin{equation}
I_\theta = -4\pi\int_0^{\frac{\pi}{2}}v_1e^{v_2+v_4+(v_3-v_4)\cos^2\theta}I_0(v_5\cos\theta)\cos\theta(1-\cos^2\theta)d\cos\theta,
\end{equation}

Setting $x=\cos\theta$, our integral becomes:
\begin{equation}
\begin{aligned}
I_\theta &= 4\pi\int_0^{1}v_1e^{v_2+v_4+(v_3-v_4)x^2}I_0(v_5x)x(1-x^2)dx\\
&= 4\pi v_1e^{v_2+v_4}\int_0^{1}e^{(v_3-v_4)x^2}I_0(v_5x)x(1-x^2)dx
\end{aligned}
\end{equation}

We find the Taylor expansion of $e^{(v_3-v_4)x^2}$:
\begin{equation}
    e^{(v_3-v_4)x^2} = \sum_{k=0}^\infty\frac{\left(v_3-v_4\right)^kx^{2k}}{k!}
\end{equation}

Now, we expand the Bessel function in series. We find that:
\begin{equation}
\begin{aligned}
    I_0(v_5x) &= \sum_{m=0}^\infty\frac{1}{m!\Gamma(m+1)}\left(\frac{v_5x}{2}\right)^{2m}
\end{aligned}
\end{equation}

Putting everything together, we arrive at:
\begin{equation}
\begin{aligned}
    I_\theta &= 4\pi v_1e^{v_2+v_4}\int_0^{1}\left(\sum_{k=0}^\infty\frac{\left(v_3-v_4\right)^kx^{2k}}{k!}\right)\left(\sum_{k=0}^\infty\frac{1}{k!\Gamma(k+1)}\left(\frac{v_5x}{2}\right)^{2k}\right)x(1-x^2)dx\\
    &= 4\pi v_1e^{v_2+v_4}\int_0^{1}\sum_{k=0}^\infty\sum_{m=0}^k\left(\frac{\left(v_3-v_4\right)^m}{m!}\right)\left(\frac{v_5^{2(k-m)}}{2^{2(k-m)}(k-m)!\Gamma(k-m+1)}\right)x^{2k+1}(1-x^2)dx\\
    &= 4\pi v_1e^{v_2+v_4}\sum_{k=0}^\infty\sum_{m=0}^k\frac{1}{2(k+1)(k+2)}\left(\frac{\left(v_3-v_4\right)^m}{m!}\right)\left(\frac{v_5^{2(k-m)}}{2^{2(k-m)}(k-m)!\Gamma(k-m+1)}\right)\\
\end{aligned}
\end{equation}

We now seek to integrate over $\beta$. Before we start, we first examine the combinations $v_2+v_4$ and $v_3-v_4$:
\begin{equation}
\begin{aligned}
    v_2 + v_4 &= (a_2\bar{a_2} + b_2\bar{b_2})\cos^2\beta\cos^2 b + (\bar{a_1}a_2 + \bar{b_1}b_2)\cos^2 b\sin^2\beta\\ 
    &+ (a_1\bar{a_2} + b_1\bar{b_2})\cos^2\beta\sin^2 b + (a_1\bar{a_1} + b_1\bar{b_1})\sin^2\beta\sin^2 b\\
    &+ (\bar{a_1}a_3 + \bar{b_1}b_3)\cos^2\beta + (a_1\bar{a_3} + b_1\bar{b_3})\cos^2 b\\
    &+ (\bar{a_2}a_3 + \bar{b_2}b_3)\sin^2\beta + (a_2\bar{a_3} + b_2\bar{b_3})\sin^2 b\\
    &= \left((a_2\bar{a_2} + b_2\bar{b_2})\cos^2 b + (a_1\bar{a_2} + b_1\bar{b_2})\sin^2 b + \bar{a_1}a_3 + \bar{b_1}b_3\right)\cos^2\beta\\
    &+ \left((\bar{a_1}a_2 + \bar{b_1}b_2)\cos^2 b + (a_1\bar{a_1} + b_1\bar{b_1})\sin^2 b + \bar{a_2}a_3 + \bar{b_2}b_3\right)\sin^2\beta\\
    &+ (a_1\bar{a_3} + b_1\bar{b_3})\cos^2 b + (a_2\bar{a_3} + b_2\bar{b_3})\sin^2 b
\end{aligned}
\end{equation}

We note that $v_1 = \sin2\beta\sin2b$, which means we can repeat the process of rewriting $\sin2\beta$ as $2\sin\beta\cos\beta$, but absorbing $\cos\beta$ behind the derivative instead. Then we rewrite the expression above as:

\begin{equation}
\begin{aligned}
    v_2 + v_4 &= \left((a_2\bar{a_2} + b_2\bar{b_2})\cos^2 b + (a_1\bar{a_2} + b_1\bar{b_2})\sin^2 b + \bar{a_1}a_3 + \bar{b_1}b_3\right)\left(1-\sin^2\beta\right)\\
    &+ \left((\bar{a_1}a_2 + \bar{b_1}b_2)\cos^2 b + (a_1\bar{a_1} + b_1\bar{b_1})\sin^2 b + \bar{a_2}a_3 + \bar{b_2}b_3\right)\sin^2\beta\\
    &+ (a_1\bar{a_3} + b_1\bar{b_3})\cos^2 b + (a_2\bar{a_3} + b_2\bar{b_3})\sin^2 b\\
    &= \bar{a_1}a_3 + \bar{b_1}b_3 + \left(a_1\bar{a_3} + b_1\bar{b_3} + a_2\bar{a_2} + b_2\bar{b_2}\right)\cos^2 b + \left(a_2\bar{a_3} + b_2\bar{b_3} + a_1\bar{a_2} + b_1\bar{b_2}\right)\sin^2 b\\
    &+ \left(\bar{a_2}a_3 + \bar{b_2}b_3 + (\bar{a_1}a_2 + \bar{b_1}b_2 - a_2\bar{a_2} - b_2\bar{b_2})\cos^2 b + (a_1\bar{a_1} + b_1\bar{b_1} - a_1\bar{a_2} - b_1\bar{b_2})\sin^2 b\right)\sin^2\beta
\end{aligned}
\end{equation}

We then turn to $v_3-v_4$:
\begin{equation}
\begin{aligned}
    v_3 - v_4 &= (a_3\bar{a_3} + b_3\bar{b_3}) + (a_1\bar{a_1} + b_1\bar{b_1})\cos^2\beta\cos^2 b + (a_1\bar{a_2} + b_1\bar{b_2})\cos^2 b\sin^2\beta\\
    &+ (\bar{a_1}a_2 + \bar{b_1}b_2)\cos^2\beta\sin^2 b + (a_2\bar{a_2} + b_2\bar{b_2})\sin^2\beta\sin^2 b\\
    &-(\bar{a_1}a_3 + \bar{b_1}b_3)\cos^2\beta - (a_1\bar{a_3} + b_1\bar{b_3})\cos^2 b\\
    &- (\bar{a_2}a_3 + \bar{b_2}b_3)\sin^2\beta - (a_2\bar{a_3} + b_2\bar{b_3})\sin^2 b\\
    &= (a_3\bar{a_3} + b_3\bar{b_3}) - (a_1\bar{a_3} + b_1\bar{b_3})\cos^2 b - (a_2\bar{a_3} + b_2\bar{b_3})\sin^2 b\\
    &+\left((a_1\bar{a_1} + b_1\bar{b_1})\cos^2 b + (\bar{a_1}a_2 + \bar{b_1}b_2)\sin^2 b -(\bar{a_1}a_3 + \bar{b_1}b_3)\right)\left(1-\sin^2\beta\right)\\
    &+ \left((a_1\bar{a_2} + b_1\bar{b_2})\cos^2 b + (a_2\bar{a_2} + b_2\bar{b_2})\sin^2 b - (\bar{a_2}a_3 + \bar{b_2}b_3)\right)\sin^2\beta\\
    &= a_3\bar{a_3} + b_3\bar{b_3} - \bar{a_1}a_3 - \bar{b_1}b_3 + (a_1\bar{a_1} + b_1\bar{b_1} - a_1\bar{a_3} - b_1\bar{b_3})\cos^2 b + (\bar{a_1}a_2 + \bar{b_1}b_2 - a_2\bar{a_3} - b_2\bar{b_3})\sin^2 b\\
    &+ \left((a_1\bar{a_2} + b_1\bar{b_2} - a_1\bar{a_1} - b_1\bar{b_1})\cos^2 b + (a_2\bar{a_2} + b_2\bar{b_2} - \bar{a_1}a_2 - \bar{b_1}b_2)\sin^2 b\right)\sin^2\beta\\
    &+ \left(\bar{a_1}a_3 + \bar{b_1}b_3 - \bar{a_2}a_3 - \bar{b_2}b_3\right)\sin^2\beta
\end{aligned}
\end{equation}

We once again regroup and relabel our coefficients for ease of computation:

\begin{equation}
\begin{aligned}
    u_1 &= \bar{a_1}a_3 + \bar{b_1}b_3 + (a_1\bar{a_3} + b_1\bar{b_3} + a_2\bar{a_2} + b_2\bar{b_2})\cos^2 b + (a_2\bar{a_3} + b_2\bar{b_3} + a_1\bar{a_2} + b_1\bar{b_2})\sin^2 b\\
    u_2 &= \bar{a_2}a_3 + \bar{b_2}b_3 + (\bar{a_1}a_2 + \bar{b_1}b_2 - a_2\bar{a_2} - b_2\bar{b_2})\cos^2 b + (a_1\bar{a_1} + b_1\bar{b_1} - a_1\bar{a_2} - b_1\bar{b_2})\sin^2 b\\
    u_3 &= a_3\bar{a_3} + b_3\bar{b_3} - \bar{a_1}a_3 - \bar{b_1}b_3 + (a_1\bar{a_1} + b_1\bar{b_1} - a_1\bar{a_3} - b_1\bar{b_3})\cos^2 b + (\bar{a_1}a_2 + \bar{b_1}b_2 - a_2\bar{a_3} - b_2\bar{b_3})\sin^2 b\\
    u_4 &= \bar{a_1}a_3 + \bar{b_1}b_3 - \bar{a_2}a_3 - \bar{b_2}b_3 + (a_1\bar{a_2} + b_1\bar{b_2} - a_1\bar{a_1} - b_1\bar{b_1})\cos^2 b + (a_2\bar{a_2} + b_2\bar{b_2} - \bar{a_1}a_2 - \bar{b_1}b_2)\sin^2 b\\
    u_5 &= (a_1\bar{a_1} - \bar{a_1}a_2 - a_1\bar{a_2} + a_2\bar{a_2} + b_1\bar{b_1} - \bar{b_1}b_2 - b_1\bar{b_2} + b_2\bar{b_2})\sin2b
\end{aligned}
\end{equation}

We set $y = \sin\beta$. Then our integral over $\beta$ becomes:

\begin{equation}
\begin{aligned}
    I_\beta &= 8\pi \sin2b \int_0^1 e^{u_1+u_2y}\sum_{k=0}^\infty\sum_{m=0}^k\frac{1}{2(k+1)(k+2)}\left(\frac{\left(u_3+u_4y^2\right)^m}{m!}\right)\left(\frac{\left(u_5y\sqrt{1-y^2}\right)^{2(k-m)}}{2^{2(k-m)}(k-m)!\Gamma(k-m+1)}\right)ydy\\
\end{aligned}
\end{equation}

We now examine $\left(u_3+u_4y^2\right)^m$. We know that we can use the binomial expansion to express it as:
\begin{equation}
    \left(u_3+u_4y^2\right)^m = \sum_{j=0}^m\binom{m}{j}u_3^j\left(u_4y^2\right)^{m-j}
\end{equation}

Then we have:
\begin{equation}
\frac{\left(u_3+u_4y^2\right)^m}{m!} = \sum_{j=0}^m\frac{1}{(m-j)!j!}u_3^j\left(u_4y^2\right)^{m-j}
\end{equation}

We then examine $\left(y\sqrt{1-y^2}\right)^{2(k-m)}$. First, we note that we can rewrite this expression as $\left(y^2-y^4\right)^{k-m}$. Then, using the binomial series, we find:
\begin{equation}
\begin{aligned}
    \left(y^2-y^4\right)^{k-m} &= \sum_{l=0}^{k-m}\binom{k-m}{l}(-1)^{k-m-l}y^{2l}y^{4(k-m-l)}\\
    &= \sum_{l=0}^{k-m}\binom{k-m}{l}(-1)^{k-m-l}y^{4k-4m-2l}
\end{aligned}
\end{equation}

Then we have:
\begin{equation}
\begin{aligned}
    \frac{\left(u_5y\sqrt{1-y^2}\right)^{2(k-m)}}{2^{2(k-m)}(k-m)!\Gamma(k-m+1)} &= \frac{u_5^{2(k-m)}\sum_{l=0}^{k-m}\binom{k-m}{l}(-1)^{k-m-l}y^{4k-4m-2l}}{2^{2(k-m)}(k-m)!\Gamma(k-m+1)}\\
    &= \sum_{l=0}^{k-m}\frac{(-1)^{k-m-l}u_5^{2(k-m)}y^{4k-4m-2l}}{2^{2(k-m)}l!(k-m-l)!\Gamma(k-m+1)}
\end{aligned}
\end{equation}

We now expand:
\begin{equation}
\begin{aligned}
    &\sum_{k=0}^\infty\sum_{m=0}^k\frac{1}{2(k+1)(k+2)}\left(\sum_{j=0}^m\frac{u_3^ju_4^{m-j}}{(m-j)!j!}y^{2(m-j)}\right)\left(\sum_{l=0}^{k-m}\frac{(-1)^{k-m-l}u_5^{2(k-m)}y^{4k-4m-2l}}{2^{2(k-m)}l!(k-m-l)!\Gamma(k-m+1)}\right)\\
    &=\sum_{k=0}^\infty\sum_{m=0}^k\frac{y^{4k-2m}}{2(k+1)(k+2)}\left(\sum_{j=0}^m\frac{u_3^ju_4^{m-j}}{(m-j)!j!}y^{-2j}\right)\left(\sum_{l=0}^{k-m}\frac{(-1)^{k-m-l}u_5^{2(k-m)}y^{-2l}}{2^{2(k-m)}l!(k-m-l)!\Gamma(k-m+1)}\right)\\
    &=\sum_{k=0}^\infty\sum_{m=0}^k\sum_{n=0}^{k}\sum_{\substack{j+l=n\\0\leq j\leq m\\0\leq l \leq k-m}}\frac{1}{2(k+1)(k+2)}\frac{u_3^ju_4^{m-j}}{(m-j)!j!}\frac{(-1)^{k-m-l}u_5^{2(k-m)}}{2^{2(k-m)}(l)!(k-m-l)!\Gamma(k-m+1)}y^{4k-2m-2n}
\end{aligned}
\end{equation}

Finally, we note that we can expand $e^{u_2y}$ using the Taylor series:
\begin{equation}
    e^{u_2y} = \sum_{p=0}^\infty\frac{u_2^py^p}{p!}
\end{equation}

Then $I_\beta$ evaluates to:
\begin{equation}
\begin{aligned}
     &8\pi \sin2b e^{u_1}\int_0^1 \sum_{p=0}^\infty\frac{u_2^py^p}{p!}\\
     &\sum_{k=0}^\infty\sum_{m=0}^k\sum_{n=0}^{k}\sum_{\substack{j+l=n\\0\leq j\leq m\\0\leq l \leq k-m}}\frac{1}{2(k+1)(k+2)}\frac{u_3^ju_4^{m-j}}{(m-j)!j!}\frac{(-1)^{k-m-l}u_5^{2(k-m)}}{2^{2(k-m)}(l)!(k-m-l)!\Gamma(k-m+1)}y^{4k-2m-2n+1}dy\\
     &=8\pi \sin2b e^{u_1}\sum_{q=0}^\infty\sum_{\substack{p+4k=q\\p,k\geq0}}\sum_{m=0}^k\sum_{n=0}^{k}\sum_{\substack{j+l=n\\0\leq j\leq m\\0\leq l \leq k-m}}\frac{u_2^p}{(k+1)(k+2)p!(q-2m-2n+2)}\frac{u_3^ju_4^{m-j}}{(m-j)!j!}\\
     &\times\frac{(-1)^{k-m-l}u_5^{2(k-m)}}{2^{2(k-m)+1}(l)!(k-m-l)!\Gamma(k-m+1)}\\
\end{aligned}
\end{equation}

Finally, we integrate over $b$. As before, we take $\sin2b$ and rewrite it as $2\sin b\cos b$. Then we absorb $\cos b$ behind the derivative and set $z = \sin b$. We now integrate over $z$ from $0$ to $1$. We rewrite $u_i$ to reflect this change:
\begin{equation}
\begin{aligned}
    u_1 &= \bar{a_1}a_3 + \bar{b_1}b_3 + a_1\bar{a_3} + b_1\bar{b_3} + a_2\bar{a_2} + b_2\bar{b_2}\\
    &+ (a_2\bar{a_3} + b_2\bar{b_3} + a_1\bar{a_2} + b_1\bar{b_2} -a_1\bar{a_3} - b_1\bar{b_3} - a_2\bar{a_2} - b_2\bar{b_2})\sin^2 b\\
    u_2 &= \bar{a_2}a_3 + \bar{b_2}b_3 + \bar{a_1}a_2 + \bar{b_1}b_2 - a_2\bar{a_2} - b_2\bar{b_2}\\
    &+ (a_1\bar{a_1} + b_1\bar{b_1} - a_1\bar{a_2} - b_1\bar{b_2} - \bar{a_1}a_2 - \bar{b_1}b_2 + a_2\bar{a_2} + b_2\bar{b_2})\sin^2 b\\
    u_3 &= a_3\bar{a_3} + b_3\bar{b_3} - \bar{a_1}a_3 - \bar{b_1}b_3 + a_1\bar{a_1} + b_1\bar{b_1} - a_1\bar{a_3} - b_1\bar{b_3}\\
    &+ (\bar{a_1}a_2 + \bar{b_1}b_2 - a_2\bar{a_3} - b_2\bar{b_3} - a_1\bar{a_1} - b_1\bar{b_1} + a_1\bar{a_3} + b_1\bar{b_3})\sin^2 b\\
    u_4 &= \bar{a_1}a_3 + \bar{b_1}b_3 - \bar{a_2}a_3 - \bar{b_2}b_3 + a_1\bar{a_2} + b_1\bar{b_2} - a_1\bar{a_1} - b_1\bar{b_1}\\
    &+ (a_2\bar{a_2} + b_2\bar{b_2} - \bar{a_1}a_2 - \bar{b_1}b_2 - a_1\bar{a_2} - b_1\bar{b_2} + a_1\bar{a_1} + b_1\bar{b_1})\sin^2 b\\
    u_5 &= (a_1\bar{a_1} - \bar{a_1}a_2 - a_1\bar{a_2} + a_2\bar{a_2} + b_1\bar{b_1} - \bar{b_1}b_2 - b_1\bar{b_2} + b_2\bar{b_2})\sin2b
\end{aligned}
\end{equation}

We set:
\begin{equation}
\label{eqn:wexpressions}
\begin{aligned}
    w_1 &= \bar{a_1}a_3 + \bar{b_1}b_3 + a_1\bar{a_3} + b_1\bar{b_3} + a_2\bar{a_2} + b_2\bar{b_2}\\
    w_2 &= a_2\bar{a_3} + b_2\bar{b_3} + a_1\bar{a_2} + b_1\bar{b_2} -a_1\bar{a_3} - b_1\bar{b_3} - a_2\bar{a_2} - b_2\bar{b_2}\\
    w_3 &= \bar{a_2}a_3 + \bar{b_2}b_3 + \bar{a_1}a_2 + \bar{b_1}b_2 - a_2\bar{a_2} - b_2\bar{b_2}\\
    w_4 &= a_1\bar{a_1} + b_1\bar{b_1} - a_1\bar{a_2} - b_1\bar{b_2} - \bar{a_1}a_2 - \bar{b_1}b_2 + a_2\bar{a_2} + b_2\bar{b_2}\\
    w_5 &= a_3\bar{a_3} + b_3\bar{b_3} - \bar{a_1}a_3 - \bar{b_1}b_3 + a_1\bar{a_1} + b_1\bar{b_1} - a_1\bar{a_3} - b_1\bar{b_3}\\
    w_6 &= \bar{a_1}a_2 + \bar{b_1}b_2 - a_2\bar{a_3} - b_2\bar{b_3} - a_1\bar{a_1} - b_1\bar{b_1} + a_1\bar{a_3} + b_1\bar{b_3}\\
    w_7 &= \bar{a_1}a_3 + \bar{b_1}b_3 - \bar{a_2}a_3 - \bar{b_2}b_3 + a_1\bar{a_2} + b_1\bar{b_2} - a_1\bar{a_1} - b_1\bar{b_1}\\
    w_8 &= a_2\bar{a_2} + b_2\bar{b_2} - \bar{a_1}a_2 - \bar{b_1}b_2 - a_1\bar{a_2} - b_1\bar{b_2} + a_1\bar{a_1} + b_1\bar{b_1}\\
    w_9 &= 2(a_1\bar{a_1} - \bar{a_1}a_2 - a_1\bar{a_2} + a_2\bar{a_2} + b_1\bar{b_1} - \bar{b_1}b_2 - b_1\bar{b_2} + b_2\bar{b_2})
\end{aligned}
\end{equation}

Then our integral becomes:
\begin{equation}
\begin{aligned}
    I_b &= 16\pi e^{w_1}\int_0^1 e^{w_2z^2}\sum_{q=0}^\infty\sum_{\substack{p+4k=q\\p,k\geq0}}\sum_{m=0}^k\sum_{n=0}^{k}\sum_{\substack{j+l=n\\0\leq j\leq m\\0\leq l \leq k-m}}\frac{(w_3+w_4z^2)^p(w_5+w_6z^2)^j(w_7+w_8z^2)^{m-j}}{(k+1)(k+2)(q-2m-2n+2)p!j!(m-j)!}\\
     &\times\frac{(-1)^{k-m-l}(w_9z\sqrt{1-z^2})^{2(k-m)}}{2^{2(k-m)+1}l!(k-m-l)!\Gamma(k-m+1)}zdz
\end{aligned}
\end{equation}

As before, we note that:
\begin{equation}
    \frac{\left(w_3+w_4z^2\right)^p}{p!} = \sum_{r=0}^p\frac{1}{(p-r)!r!}w_3^r\left(w_4z^2\right)^{p-r}
\end{equation}
and
\begin{equation}
\frac{\left(w_5+w_6z^2\right)^j}{j!} = \sum_{s=0}^j\frac{1}{(j-s)!s!}w_5^s\left(w_6z^2\right)^{j-s}
\end{equation}
and
\begin{equation}   \frac{\left(w_7+w_8z^2\right)^{m-j}}{(m-j)!} = \sum_{t=0}^{m-j}\frac{1}{(m-j-t)!t!}w_7^t\left(w_8z^2\right)^{m-j-t}
\end{equation}

We find that:
\begin{equation}
\begin{aligned}
\frac{(w_3+w_4z^2)^p(w_5+w_6z^2)^j(w_7+w_8z^2)^{m-j}}{p!j!(m-j)!} &= \left(\sum_{r=0}^p\frac{w_3^rw_4^{p-r}}{(p-r)!r!}z^{2p-2r}\right)\left(\sum_{s=0}^j\frac{w_5^sw_6^{j-s}}{(j-s)!s!}z^{2j-2s}\right)\\
&\times\left(\sum_{t=0}^{m-j}\frac{w_7^tw_8^{m-j-t}}{(m-j-t)!t!}z^{2m-2j-2t}\right)\\
&= \sum_{g=0}^{p+m}\sum_{\substack{r+s+t=g\\0\leq r\leq p\\0\leq s\leq j\\0\leq t\leq m-j}}\frac{w_3^rw_4^{p-r}w_5^sw_6^{j-s}w_7^tw_8^{m-j-t}}{(p-r)!r!(j-s)!s!(m-j-t)!t!}z^{2(p+m-g)}
\end{aligned}
\end{equation}

We also know that $\left(z\sqrt{1-z^2}\right)^{2(k-m)}$ can be written as:
\begin{equation}
    \left(z\sqrt{1-z^2}\right)^{2(k-m)} = \sum_{h=0}^{k-m}\binom{k-m}{h}(-1)^{k-m-h}z^{4k-4m-2h}
\end{equation}

Then we have:
\begin{equation}
\begin{aligned}
    \frac{(-1)^{k-m-l}(w_9z\sqrt{1-z^2})^{2(k-m)}}{2^{2(k-m)+1}l!(k-m-l)!\Gamma(k-m+1)} &= \frac{(-1)^{k-m-l}w_9^{2(k-m)}\sum_{h=0}^{k-m}\binom{k-m}{h}(-1)^{k-m-h}z^{4k-4m-2h}}{2^{2(k-m)+1}l!(k-m-l)!\Gamma(k-m+1)}\\
    &= \sum_{h=0}^{k-m}\frac{(-1)^{-l-h}w_9^{2(k-m)}z^{4k-4m-2h}}{2^{2(k-m)+1}l!h!(k-m-l)!(k-m-h)!}
\end{aligned}
\end{equation}

We compute:
\begin{equation*}
\begin{aligned}
    &\sum_{q=0}^\infty\sum_{\substack{p+4k=q\\p,k\geq0}}\sum_{m=0}^k\sum_{n=0}^{k}\sum_{\substack{j+l=n\\0\leq j\leq m\\0\leq l \leq k-m}}\frac{1}{(k+1)(k+2)(q-2m-2n+2)}\\
    &\times\left(\sum_{g=0}^{p+m}\sum_{\substack{r+s+t=g\\0\leq r\leq p\\0\leq s\leq j\\0\leq t\leq m-j}}\frac{w_3^rw_4^{p-r}w_5^sw_6^{j-s}w_7^tw_8^{m-j-t}}{(p-r)!r!(j-s)!s!(m-j-t)!t!}z^{2(p+m-g)}\right)\\
    &\times\left(\sum_{h=0}^{k-m}\frac{(-1)^{-l-h}w_9^{2(k-m)}z^{4k-4m-2h}}{2^{2(k-m)+1}l!h!(k-m-l)!(k-m-h)!}\right)\\
\end{aligned}
\end{equation*}
and arrive at:
\begin{equation*}
\begin{aligned}
    &\sum_{q=0}^\infty\sum_{\substack{p+4k=q\\p,k\geq0}}\sum_{m=0}^k\sum_{n=0}^{k}\sum_{\substack{j+l=n\\0\leq j\leq m\\0\leq l \leq k-m}}\frac{1}{(k+1)(k+2)(q-2m-2n+2)}\\
    &\times\sum_{g=0}^{p+k}\sum_{\substack{r+s+t+h=g\\0\leq r\leq p\\0\leq s\leq j\\0\leq t\leq m-j\\0\leq h\leq k-m}}\frac{(-1)^{-l-h}w_3^rw_4^{p-r}w_5^sw_6^{j-s}w_7^tw_8^{m-j-t}w_9^{2(k-m)}}{2^{2(k-m)+1}(p-r)!r!(j-s)!s!(m-j-t)!t!l!h!(k-m-l)!(k-m-h)!}z^{2(2k+p-m-g)}
\end{aligned}
\end{equation*}

Once again, we expand $e^{w_2z}$ using the Taylor series:
\begin{equation}
    e^{w_2z} = \sum_{f=0}^\infty\frac{w_2^fz^f}{f!}
\end{equation}

Then our integral becomes:
\begin{equation*}
\begin{aligned}
    &16\pi e^{w_1}\int_0^1 \left(\sum_{f=0}^\infty\frac{w_2^fz^f}{f!}\right)\sum_{q=0}^\infty\sum_{\substack{p+4k=q\\p,k\geq0}}\sum_{m=0}^k\sum_{n=0}^{k}\sum_{\substack{j+l=n\\0\leq j\leq m\\0\leq l \leq k-m}}\frac{1}{(k+1)(k+2)(q-2m-2n+2)}\\
    &\times\sum_{g=0}^{p+k}\sum_{\substack{r+s+t+h=g\\0\leq r\leq p\\0\leq s\leq j\\0\leq t\leq m-j\\0\leq h\leq k-m}}\frac{(-1)^{-l-h}w_3^rw_4^{p-r}w_5^sw_6^{j-s}w_7^tw_8^{m-j-t}w_9^{2(k-m)}}{2^{2(k-m)+1}(p-r)!r!(j-s)!s!(m-j-t)!t!l!h!(k-m-l)!(k-m-h)!}z^{2(2k+p-m-g)}zdz\\
\end{aligned}
\end{equation*}

This is a hideous series, and we would be forgiven for thinking that we should define a new index that matches $2(2k+p-m-g)$. But $q$ does the job, if more subtly, and so we will retain $q$ and rewrite $2(2k+p-m-g)$ as $2(q-2k-m-g)$. Then we can integrate over $z$ and find:
\begin{equation}
\begin{aligned}
    &16\pi e^{w_1}\int_0^1\sum_{d=0}^\infty\sum_{\substack{2q-4k+f=d\\p+4k=q\\q,k,p,f\geq0}}\sum_{m=0}^k\sum_{n=0}^{k}\sum_{\substack{j+l=n\\0\leq j\leq m\\0\leq l \leq k-m}}\frac{1}{(k+1)(k+2)(q-2m-2n+2)}\\
    &\times\sum_{g=0}^{p+k}\sum_{\substack{r+s+t+h=g\\0\leq r\leq p\\0\leq s\leq j\\0\leq t\leq m-j\\0\leq h\leq k-m}}\frac{(-1)^{-l-h}w_2^fw_3^rw_4^{p-r}w_5^sw_6^{j-s}w_7^tw_8^{m-j-t}w_9^{2(k-m)}}{2^{2(k-m)+1}f!(p-r)!r!(j-s)!s!(m-j-t)!t!l!h!(k-m-l)!(k-m-h)!}z^{d-2m-2g+1}dz\\
    &= 16\pi e^{w_1}\sum_{q=0}^\infty\sum_{\substack{2p+4k+f=q\\k,p,f\geq0}}\sum_{m,n=0}^k\sum_{\substack{j+l=n\\0\leq j\leq m\\0\leq l \leq k-m}}\sum_{g=0}^{p+k}\sum_{\substack{r+s+t+h=g\\0\leq r\leq p\\0\leq s\leq j\\0\leq t\leq m-j\\0\leq h\leq k-m}}\frac{1}{(k+1)(k+2)(p+4k-2m-2n+2)}\\
    &\times\frac{(-1)^{-l-h}w_2^fw_3^rw_4^{p-r}w_5^sw_6^{j-s}w_7^tw_8^{m-j-t}w_9^{2(k-m)}}{2^{2(k-m)+1}(q-2m-2g+2)f!(p-r)!r!(j-s)!s!(m-j-t)!t!l!h!(k-m-l)!(k-m-h)!}
\end{aligned}
\end{equation}

For clarity, we rewrite our integral as:
\begin{equation}
\label{eqn:su3formula}
\begin{aligned}
    I &= \sum_{q=0}^\infty\sum_{\substack{f+2p+4k=q\\f,p,k\geq0}}\sum_{m,n=0}^k\sum_{\substack{j+l=n\\0\leq j\leq m\\0\leq l \leq k-m}}\sum_{g=0}^{p+k}\sum_{\substack{r+s+t+h=g\\0\leq r\leq p\\0\leq s\leq j\\0\leq t\leq m-j\\0\leq h\leq k-m}}\frac{16\pi e^{w_1}}{(k+1)(k+2)(p+4k-2m-2n+2)(q-2m-2g+2)}\\
    &\times\frac{(-1)^{-l-h}w_2^fw_3^rw_4^{p-r}w_5^sw_6^{j-s}w_7^tw_8^{m-j-t}w_9^{2(k-m)}}{2^{2(k-m)+1}f!(p-r)!r!(j-s)!s!(m-j-t)!t!l!h!(k-m-l)!(k-m-h)!}\\
\end{aligned}
\end{equation}

We now seek to simplify this answer. First, starting from $q=0$, we list the first few possible combinations for $q$, $f$, $p$, and $k$ for the first term. Our table is as follows:

\begin{center}
\begin{tabular}{ cccc }
\begin{tabular}{||c c c c||} 
 \hline
 q & f & p & k \\ [0.5ex] 
 \hline\hline
 0 & 0 & 0 & 0 \\ 
 \hline
 1 & 1 & 0 & 0 \\
 \hline
 2 & 0 & 1 & 0 \\
 \hline
 2 & 2 & 0 & 0 \\
 \hline
 3 & 1 & 1 & 0 \\
 \hline
 3 & 3 & 0 & 0 \\ 
 \hline
 4 & 0 & 0 & 1 \\
 \hline
 4 & 0 & 2 & 0 \\
 \hline
 4 & 2 & 1 & 0 \\
 \hline
\end{tabular}
\begin{tabular}{||c c c c||} 
 \hline
 q & f & p & k \\ [0.5ex] 
 \hline\hline
 4 & 4 & 0 & 0 \\
 \hline
 5 & 1 & 0 & 1 \\
 \hline
 5 & 1 & 2 & 0 \\ 
 \hline
 5 & 3 & 1 & 0 \\
 \hline
 5 & 5 & 0 & 0 \\
 \hline
 6 & 0 & 1 & 1 \\
 \hline
 6 & 0 & 3 & 0 \\
 \hline
 6 & 2 & 2 & 0 \\ 
 \hline
 6 & 4 & 1 & 0 \\
 \hline
\end{tabular}
\begin{tabular}{||c c c c||} 
 \hline
 q & f & p & k \\ [0.5ex] 
 \hline\hline
 6 & 6 & 0 & 0 \\
 \hline
 7 & 1 & 1 & 1 \\
 \hline
 7 & 1 & 3 & 0 \\
 \hline
 7 & 3 & 2 & 0 \\ 
 \hline
 7 & 5 & 1 & 0 \\
 \hline
 7 & 7 & 0 & 0 \\
 \hline
 8 & 0 & 0 & 2 \\
 \hline
 8 & 0 & 2 & 1 \\ 
 \hline
 8 & 0 & 4 & 0 \\
 \hline
\end{tabular}
\begin{tabular}{||c c c c||} 
 \hline
 q & f & p & k \\ [0.5ex] 
 \hline\hline
 8 & 2 & 1 & 1 \\
 \hline
 8 & 2 & 3 & 0 \\
 \hline
 8 & 4 & 0 & 1 \\
 \hline
 8 & 4 & 2 & 0 \\ 
 \hline
 8 & 6 & 1 & 0 \\
 \hline
 8 & 8 & 0 & 0 \\
 \hline
 9 & 1 & 0 & 2 \\ 
 \hline
 9 & 1 & 2 & 1 \\
 \hline
 9 & 1 & 4 & 0 \\
 \hline
\end{tabular}
\end{tabular}
\end{center}

We see that if we hold all the other indices constant and expand the sum over $q$ and $f$, then we may extract a factor of $\frac{1}{q-2m-2g+2}\frac{w_2^f}{f!}$ from the terms associated with the set of fixed indices. Since $q-f = 2p+4k$, we may rewrite this factor as $\frac{1}{f+2p+2k-2m-2g+2}\frac{w_2^f}{f!}$. Summing from zero to infinity, we find that:
\begin{equation}
\sum_{f=0}^\infty\frac{1}{f+2p+4k-2m-2g+2}\frac{\omega_2^f}{f!} = (-w_2)^{-2p-4k+2m+2g-2}\Gamma(2p+4k-2m-2g+2, 0, -w_2)
\end{equation}

Then we have:
\begin{equation}
\begin{aligned}
    I &= \sum_{p,k\geq0}^\infty\sum_{m,n=0}^k\sum_{\substack{j+l=n\\0\leq j\leq m\\0\leq l \leq k-m}}\sum_{g=0}^{p+k}\sum_{\substack{r+s+t+h=g\\0\leq r\leq p\\0\leq s\leq j\\0\leq t\leq m-j\\0\leq h\leq k-m}}\frac{\pi e^{w_1}\Gamma(2p+4k-2m-2g+2, 0, -w_2)}{(k+1)(k+2)(p+4k-2m-2n+2)(-w_2)^{2p+4k-2m-2g+2}}\\
    &\times\frac{(-1)^{-l-h}w_3^rw_4^{p-r}w_5^sw_6^{j-s}w_7^tw_8^{m-j-t}w_9^{2(k-m)}}{2^{2(k-m)-3}(p-r)!r!(j-s)!s!(m-j-t)!t!l!h!(k-m-l)!(k-m-h)!}\\
\end{aligned}
\end{equation}

\section{Expansion in Terms of Restricted Characters}
\label{sec:appendixB}

We seek to compute our four-matrix Harish-Chandra integral through a character expansion. We start with: 
\bal
I_{R,R^\pr}=\int \d U\, \chi_R (U X U^\dagger \bar{X}) \chi_{R^\pr} (U Y U^\dagger \bar{Y})
\eal
If we have $A\in \Hom(V,V)$, $A^\pr \in \Hom(V^\pr,V^\pr)$, then:
\bal
\Tr_V A  ~ \Tr_{V^\pr} A^\pr = \Tr_{V\otimes V^\pr} A \otimes A^\pr
\eal
We arrive at:
\bal
I_{R,R^\pr}=\int \d U\, \chi_{R\otimes R^\pr}  \( \, U_R \otimes U_{R^\pr} \,  X_R\otimes Y_{R^\pr} \, U_R^\dagger \otimes U_{R^\pr}^\dagger \, \bar{X}_R \otimes \bar{X}_{R^\pr} \)
\eal
Now we may decompose our product representation into irreducible representations:
\bal
I_{R,R^\pr}=\sum_{S\in R\otimes R^\pr}\int \d U \, \chi_S \( U_S Z_S U_S^\dagger\bar{Z}_S \)
\eal
where
\bal
Z_S \in X_R \otimes Y_{R^\pr}
\eal

For example, if we set $R=
\yt{~}$, $R^\pr=
\yt{~}$, then we have:
\bal
\yt{x} \otimes
\yt{y}=\yt{x&y}
\oplus
\yt{x\\y}
\eal
\bal
\yt{x&y}=\frac{1}{2} \([x][y]+[xy]\),~~~\yt{x\\y}=\frac{1}{2} \([x][y]-[xy]\)
\eal
From here on out, we use symbols interchangeably to represent both the representation $S$ and the character $\chi_S(Z_S)$ associated with it. We write the trace of the fundamental representation matrices as $[x]\equiv \sum_{i=1}^N x_i$. We see then that $S$ could be the symmetric/anti-symmmetric combination of $X$ and $Y$ in its fundamental representation. If we seek to evaluate our integral with the Young diagrams we listed earlier, we arrive at:
\bal
I_{R,R}=\frac{1}{D_S}\chi_S (X \otimes Y)\chi_S (\bar{X} \otimes \bar{Y})
+\frac{1}{D_{S^\pr}}\chi_{S^\pr}(X \otimes Y)\chi_{S^\pr}(\bar{X} \otimes \bar{Y}),~~~S=\yt{x&y},~~~S^\pr=\yt{x\\y}
\eal
where $D_S$ is the dimension of the $R$ representation of the $GL(|R|)$ group.

\hfill \break

We now compute our second example, where we hae set $R=\yt{~&~}$, $R^\pr =\yt{~}$. Then we have:
\bal
\yt{x&x} \otimes \yt{y}= \yt{x&x&y} \oplus \yt{x&x\\y}
\eal
\bal
\yt{x&x&y}&=\frac{1}{6} \([x]^2[y]+[x^2][y]+2[x][xy]+2[x^2y]\),\nn
\yt{x&x\\y}&=\frac{1}{3} \([x]^2[y]+[x^2][y]-[x][xy]-[x^2y]\)
\eal
\footnote{
We must be careful about how we project our tensors onto the Young tableaux. We have:
\bal
&T^{ab}=T^{ba},~~~T^{ab|c}=T_S^{abc}+T_{S^\pr}^{ab,c}\nn
&T_S^{abc}=\frac{1}{3}(T^{ab|c}+T^{bc|a}+T^{ca|b}) ,~~~T_{S^\pr}^{ab,c}=\frac{1}{3}(2T^{ab|c}-T^{bc|a}-T^{ca|b})
\eal
where $T_S$ is totally symmetric as expected. However, projecting $T_{S^\pr}$ requires more delicate handling. The tensor satisfies
\bal
T_{S^\pr}^{ab,c}+T_{S^\pr}^{bc,a}+T_{S^\pr}^{ca,b}=0
\eal
}
\bal
I_{R,R^\pr}=\frac{1}{D_S} \chi_S\(X\otimes Y\) \chi_S\(\bar{X}\otimes \bar{Y}\)+\frac{1}{D_{S^\pr}} \chi_{S^\pr} \(X\otimes Y\) \chi_{S^\pr} \(\bar{X}\otimes \bar{Y}\),~~~S=\yt{x&x&y} ,~~~S^\pr=\yt{x&x\\y}
\eal

Similarly, if $R=\yt{~\\~}$, $R^\pr =\yt{~}$, then
\bal
\yt{x\\x} \otimes \yt{y}= \yt{x&y\\x} \oplus \yt{x\\x\\y}
\eal

\bal
\yt{x&y\\x}&=\frac{1}{3} \([x]^2[y]-[x^2][y]+[x][xy]-[x^2y]\)\nn
\yt{x\\x\\y}&=\frac{1}{6} \([x]^2[y]-[x^2][y]-2[x][xy]+2[x^2y]\)
\eal
\footnote{
Similarly,
\bal
&T^{ab}=-T^{ba},~~~T^{ab|c}=T^{abc}_S+T^{ab,c}_{S^\pr}\nn
&T_S^{abc}=\frac{1}{3}(T^{ab|c}+T^{bc|a}+T^{ca|b}) ,~~~T_{S^\pr}^{ab,c}=\frac{1}{3}(2T^{ab|c}-T^{bc|a}-T^{ca|b})
\eal
}

These diagrams describe what are called the {\it restricted Schur polynomials} in literature.

\hfill \break

We have dealt with each term seperately; we now combine our results for different representations to compute the initial four-matrix Harish-Chandra integral. We find:
\bal
I &= \sum_{R,R^\pr} d_R d_{R^\pr}   \int \d U \, \chi_R (U X U^\dagger \bar{X}) \chi_{R^\pr} (U Y U^\dagger \bar{Y})\nn
&= \sum_{R,R^\pr} d_R d_{R^\pr}   \sum_{S\in R\otimes R^\pr}\int \d U \, 
\Tr_S \[U_S Z_S U_S^\dagger \bar{Z}_S\]\nn
& = \sum_{R,R^\pr} d_R d_{R^\pr} \sum_{S\in R\otimes R^\pr} \frac{1}{D_S}\Tr_S (Z_S) \Tr_S (\bar{Z}_S),~~~Z=X_R \otimes Y_{R^\pr}
\eal
where $d_R$ is the dimension of $R$ representation of $S_{|R|}$ group, divided by $|R|!$.

Equivalently, the integral can also be written as:
\bal
I&=\sum_{m,n=0}^\infty \frac{1}{m!n!} \int \d U \, \( \Tr [ U X U^\dagger \bar{X} ] \)^m  \(\Tr [ U Y U^\dagger \bar{Y} ] \)^n \nn
&=\sum_{m,n=0}^\infty \frac{1}{m!n!} \sum_{S \in V^{\otimes m}\otimes V^{\otimes n}}  \frac{1}{D_S} \Tr_S (Z_S) \Tr_S (\bar{Z}_S)
\eal

The first few terms are:
\bal
I= \, 1&+ \yt{x}^2+\yt{y}^2+\yt{x&y}^2+\yt{x\\y}^2+\frac{1}{2}\yt{x&x}^2+\frac{1}{2}\yt{x\\x}^2+\frac{1}{2}\yt{y&y}^2+\frac{1}{2}\yt{y\\y}^2\nn
&+\frac{1}{3!}\(\yt{x&x&x}^2+2\yt{x&x\\x}^2+\yt{x\\x\\x}^2\)+\frac{1}{2!} \(\yt{x&x&y}^2+\yt{x&x\\y}^2+\yt{x&y\\x}^2+\yt{x\\x\\y}^2\)\nn
&+\frac{1}{2!} \(\yt{x&y&y}^2+\yt{y&y\\x}^2+\yt{x&y\\y}^2+\yt{y\\y\\x}^2\)+\frac{1}{3!}\(\yt{y&y&y}^2+2\yt{y&y\\y}^2+\yt{y\\y\\y}^2\)+...
\eal
where the square of a Young diagram represents the product of the character of $X\otimes Y$ and the character of the same representation of $\bar{X}\otimes \bar{Y}$, divided by the dimension $D_R$ of this representation, e.g.
\bal
\yt{x&y}^2 &\equiv \frac{1}{D_S}\chi_{S}(Z_S)\chi_{S}(\bar{Z}_S)\nn
&=\frac{2}{(N+1)N}\frac{[X][Y]+[XY]}{2}\frac{[\bar{X}][\bar{Y}]+[\bar{X}\bar{Y}]}{2},~~~S=\yt{x&y}
\eal

The expansion above matches \eqref{eqn:su2fixedpointformula}, the $N=2$ integral formula, precisely.

% The bibliography will probably be heavily edited during typesetting.
% We'll parse it and, using the arxiv number or the journal data, will
% query inspire, trying to verify the data (this will probalby spot
% eventual typos) and retrive the document DOI and eventual errata.
% We however suggest to always provide author, title and journal data:
% in short all the informations that clearly identify a document.

%\begin{thebibliography}{99}

% Please avoid comments such as "For a review'', "For some examples",
% "and references therein" or move them in the text. In general,
% please leave only references in the bibliography and move all
% accessory text in footnotes.

% Also, please have only one work for each \bibitem.

%\end{thebibliography}

\bibliographystyle{JHEP}
	\cleardoublepage

\renewcommand*{\bibname}{References}

\bibliography{references}
\end{document}